# Wearable-informed generative digital avatars predict task-conditioned post-stroke locomotion

## Authors

Yanning Dai[1,2], Chenyu Tang[3], Ruizhi Zhang[1], Wenyu Yang[1], Yilan Zhang[2,4], Yuhui Wang[2], Junliang Chen[1], Xuhang Chen[5], Ruimou Xie[6], Yangyue Cao[7], Qiaoying Li[8], Jin Cao[9], Tao Li[10], Hubin Zhao[11], Yu Pan[6], Arokia Nathan[12], Xin Gao[2,4], Peter Smielewski[5], Shuo Gao[1]

## Affiliations

1 School of Instrumentation and Optoelectronic Engineering, Beihang University, 100191, Beijing, China

2 Centers of Excellence for Generative AI, King Abdullah University of Science and Technology, 23955, Thuwal, Saudi Arabia

3 School of Electronic and Computer Engineering, Peking University, 518055, Shenzhen, China

4 Center of Excellence on Smart Health (KCSH), King Abdullah University of Science and Technology, 23955, Thuwal, Saudi Arabia

5 Brain Physics Laboratory, Division of Neurosurgery, Department of Clinical Neurosciences, University of Cambridge, Cambridge, CB2 0QQ, UK

6 Department of Physical Medicine and Rehabilitation, Beijing Tsinghua Changung Hospital, Tsinghua University, 100084, Beijing, China

7 Department of Rehabilitation, Beijing Tongren Hospital, Capital Medical University, 100005, Beijing, China

8 Stomatology and Rehabilitation Department, Shijiazhuang People's Hospital, 050000, Shijiazhuang, China

9 School of Life Sciences, Beijing University of Chinese Medicine, 100029, Beijing, China

10 Department of AI Medicine, The Fifth Medical Center of Chinese PLA General Hospital (301 Hospital), 100853, Beijing, China

11 Division of Surgery and Interventional Science, University College London, WC1E 6BT, London, UK

12 Darwin College, University of Cambridge, CB3 9EU, Cambridge, UK

Corresponding to: Chenyu Tang (ctang@pku.edu.cn), Yu Pan (panyu@btch.edu.cn), and Shuo Gao (shuo_gao@buaa.edu.cn)

## Abstract

Dynamic prediction of locomotor capacity after stroke could enable more individualized rehabilitation, yet current assessments largely provide static impairment scores and do not indicate whether patients can perform specific tasks such as slope walking or stair climbing. Here, we present a wearable-informed data–physics hybrid generative framework that reconstructs a stroke survivor's locomotor control from wearable inertial sensing and predicts task-conditioned post-stroke locomotion in new environments. From a single 20 m level-ground walking trial recorded by five IMUs, the framework personalizes a physics-based digital avatar using a healthy-motion prior and hybrid imitation learning, generating dynamically feasible, patient-specific movements for inclined walking and stair negotiation. Across 11 stroke inpatients, predicted postures achieved approximately 95% of the practically attainable accuracy for slope ascent and 80% for stair climbing, respectively, demonstrating reliable task-specific gait prediction. In a multicenter randomized pilot feasibility study ($n = 21$; 28 days), access to scenario-specific locomotion predictions to support task selection and difficulty titration was associated with larger gains in Fugl–Meyer lower-extremity scores than the control group (mean change 6.0 vs 3.7 points; $p < 0.05$). These results suggest that wearable-informed generative digital avatars may augment individualized gait rehabilitation planning and provide a pathway toward dynamically personalized post-stroke motor recovery strategies.

## I Main

Stroke is a leading cause of long-term motor disability and frequently results in persistent gait impairments such as hemiplegia, reduced joint mobility, and asymmetric movement patterns (1–3). These motor deficits substantially limit independence, increase fall risk, and contribute to long-term healthcare burden. Globally, approximately 11.9 million people experience a stroke each year, and more than one-third of survivors retain chronic motor dysfunction despite months of rehabilitation (4, 5). In many developing regions, disability rates can rise to 60–80 percent, reflecting not only the severity of post-stroke impairments but also the difficulty of delivering timely, intensive, and personalized rehabilitation (6, 7).

Motor rehabilitation remains the most effective pathway for restoring function (3, 8–10), but its impact depends on how well task selection and dosing are matched to an individual patient's changing capacity. Clinicians continuously adjust cadence, incline, step height, and repetition to elicit targeted neuromuscular engagement while avoiding excessive fatigue, frustration, or injury risk (11–13). To allow fine-grained quantification of gait and impairment in both laboratory and routine settings, advanced wearable sensing technologies have been introduced (14, 15). Yet most existing pipelines compress rich time-series signals into static, descriptive outputs such as standardized clinical rating scales (16–18), kinematic movement-quality indicators (15, 19–21), and predicted recovery curves or performance scores (22, 23). These summaries describe how a patient walked under an observed condition, but they fail to translate motion data into a dynamic, task-conditioned forecast of how the same individual is likely to move in a new rehabilitation scenario, which compensatory strategies will emerge, and where stability and safety limits are likely to be reached. Consequently, even when extensive wearables-generated assessments are available, therapists still rely on cautious trial-and-error when deciding, for example, whether a patient can safely progress to slope walking or stair training, which can lead to mismatched difficulty, suboptimal dosing, and delayed use of the critical rehabilitation window. The missing capability is not a new way to measure gait, but a way to convert a brief, wearable-captured gait sequence into task-level, patient-specific predictions.

To address this gap, we introduce a wearable-informed generative framework that reconstructs each patient's locomotor control from a single 20 m level-ground walking trial recorded with body-worn inertial sensors, and uses the resulting control signature to predict task-conditioned post-stroke locomotion in new scenarios. The framework instantiates a patient-specific digital avatar by coupling a proportional–derivative physics controller with a Healthy Motion Atlas that captures normative gait coordination, and a goal-conditioned deep reinforcement learning policy that synthesizes individualized, dynamically feasible movements for two common yet challenging tasks: slope ascent and stair climbing. In this formulation, a brief wearable recording provides a minimal, scalable input from which the avatar can generate locomotion beyond the observed baseline by maintaining consistency in underlying control activation patterns, while preserving patient-specific characteristics.

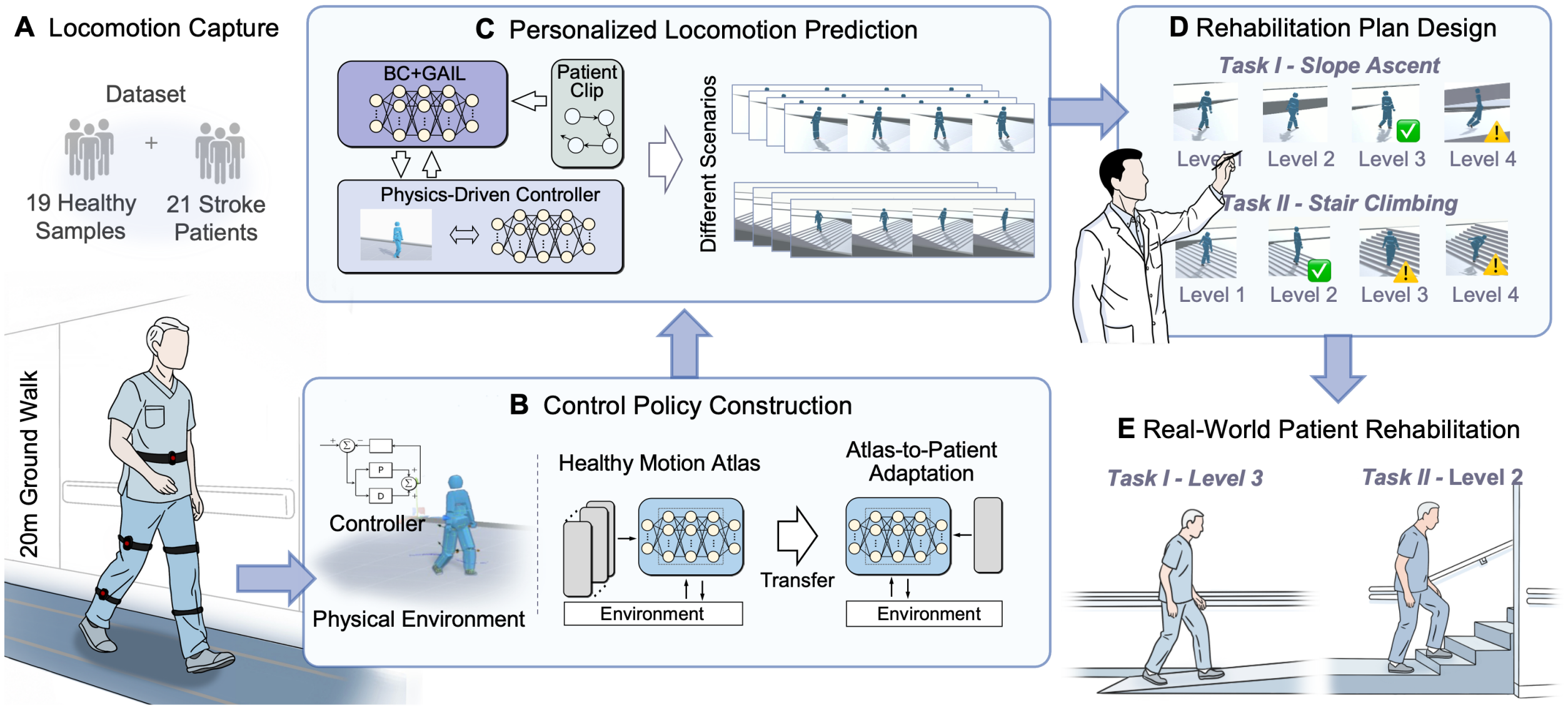

**Fig. 1. Data–physics hybrid generative framework bridging simulation and real-world rehabilitation.** Overview of the proposed generative framework, which constructs patient-specific control models from real-world walking data and generates physics-consistent locomotion predictions to support rehabilitation decision-making. (**A**) Baseline gait is recorded from a 20 m level-ground walk with wearable sensors, yielding joint kinematics, contacts, and interaction features for model initialization. (**B**) An individualized control model is constructed by distilling the Healthy Motion Atlas as a population prior, followed by Atlas-to-Patient adaptation using a patient-specific walking clip. (**C**) A data–physics hybrid generative model is built by coupling the personalized physics controller with behavioral imitation and GAIL-based adversarial learning, allowing patient-specific control patterns to predict locomotion across rehabilitation scenarios (e.g., slope, stairs). (**D**) Clinicians review the predicted locomotion videos to assess task performances, determine task-capacity limits, and select appropriate rehabilitation tasks and difficulty levels. (**E**) Patients perform the selected training, and the personalized controller can be periodically refreshed with new walking assessments, enabling its use across multiple stages of rehabilitation.

The avatar predicts task-specific locomotion after observing only a single level-ground walking trial: patient data drive personalization, whereas task training is obtained through large-scale exploration in the physics simulator, removing the need for task-specific demonstrations from patients. In a multi-center study of 21 post-stroke inpatients across five rehabilitation centers, the model generated stable task-conditioned predictions for both slope and stair scenarios and reproduced individualized locomotion patterns, with predicted motions showing high similarity to measured task kinematics. To explore potential clinical utility, the cohort was further randomized into a model-assisted group (n = 11) and a control group (n = 10) for a 28-day pilot feasibility study; therapists in the model-assisted group used the avatar's task-conditioned predictions as a reference for task selection and difficulty titration. Weekly Fugl–Meyer Assessment lower-extremity scores increased from $25.6 \pm 0.6$ to $31.6 \pm 0.6$ in the model-assisted group and from $26.0 \pm 0.4$ to $29.7 \pm 0.5$ in controls ($p < 0.05$).

Together, the results establish the proposed generative model to transform wearable gait measurements into support for scenario-specific rehabilitation decisions. By transforming a short walking clip into actionable forecasts of task performance, compensatory strategy, and task-capacity limits, the framework provides a mechanism to use wearable data for individualized planning rather than static description. The same principles hold the potential to be extended to more complex rehabilitation settings, broader neuromuscular conditions, and adaptive robotic assistance, moving toward continuously personalized and intelligent rehabilitation. Table S1 provides a comparative overview between the proposed generative model and state-of-the-art AI-assisted wearable approaches for rehabilitation.

# II. Results

### Overview of the Generative Rehabilitation Modeling Framework

We present a generative modeling framework that reconstructs each patient's control policy from a single 20 m walking clip and uses these personalized control signatures to generate task-conditioned locomotion for rehabilitation. This workflow begins with lightweight locomotion capture from healthy individuals and stroke patients (Fig. 1A). A physics-driven digital avatar is then constructed, first by learning a population-level Healthy Motion Atlas and subsequently adapting it to each patient through goal-conditioned transfer (Fig. 1B). The personalized controller generates task-conditioned locomotion across new rehabilitation scenarios (Fig. 1C), producing visualizations and activation patterns that reveal patient-specific abnormal or compensatory movement tendencies. These task-conditioned predictions support clinician-guided planning by indicating which task levels are achievable under simulation and which may induce compensations or instability (Fig. 1D), ultimately supporting real-world rehabilitation decisions by guiding task and difficulty selection (Fig. 1E). An end-to-end overview of the framework, with illustrative visualizations of key results, is provided in Supplementary Video 1.

In this study, a total of 21 post-stroke inpatients were recruited across five rehabilitation centers (Beijing Tsinghua Changgung Hospital, Beijing Tongren Hospital, Shijiazhuang People's Hospital, Beijing University of Chinese Medicine, and the Fifth Medical Center of the Chinese PLA General Hospital). Patients were randomly allocated to a model-assisted group (n = 11) or a control group (n = 10). Unless otherwise stated, all personalized modeling and quantitative evaluations reported below refer to the model-assisted group.

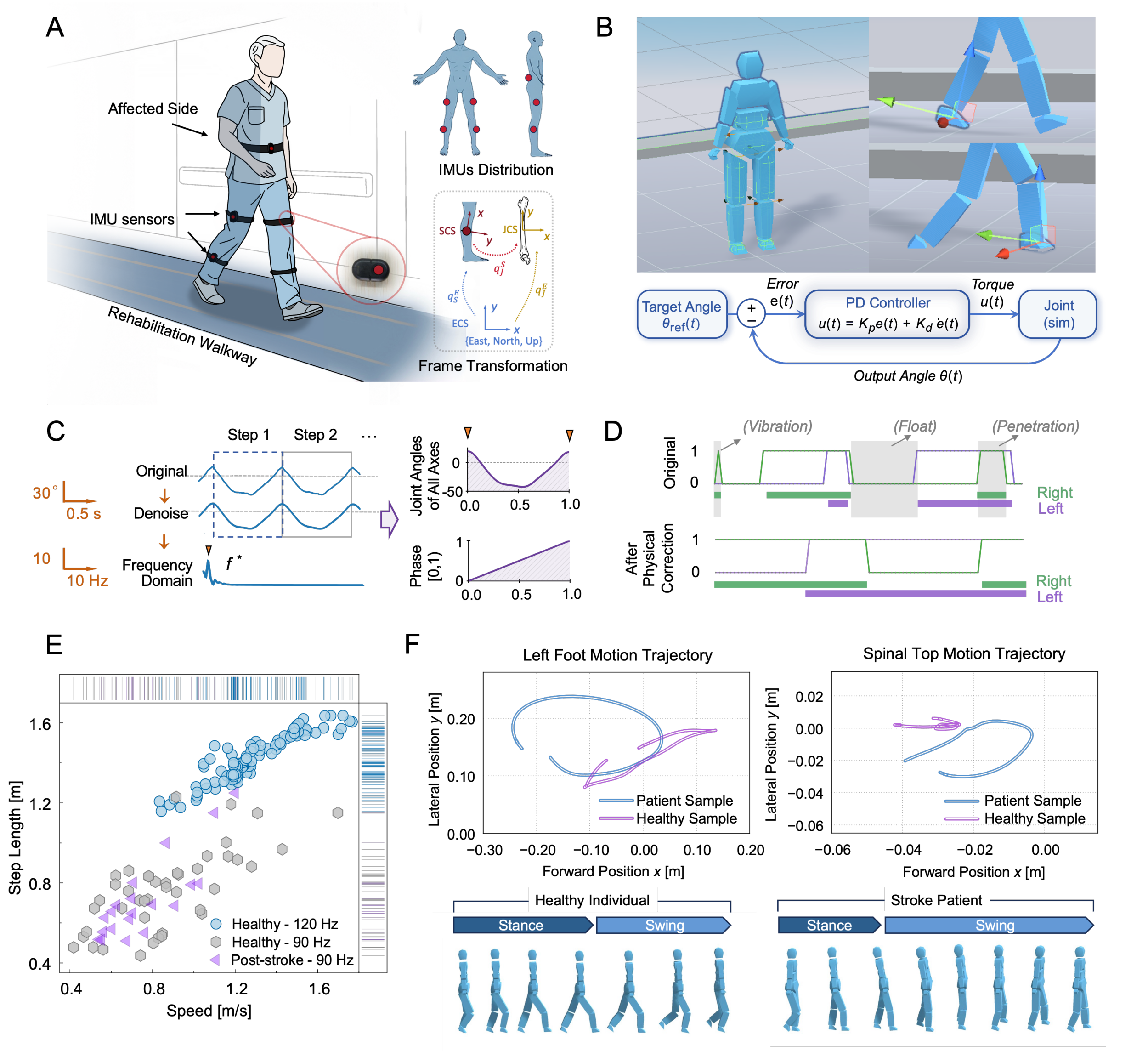

**Fig. 2. Construction of the human locomotion dataset and physics-based control model.** (**A**) Experimental setup in which a patient walks along a rehabilitation walkway wearing IMU sensors on the pelvis, bilateral thighs, and bilateral shanks. A coordinated transformation chain between the Earth Coordinate System (ECS), Segment Coordinate System (SCS), and Joint Coordinate System (JCS) ensures anatomically consistent propagation of segment rotations and translations. (**B**) Digital human avatar reconstructed from anatomical parameters, together with examples of IMU-induced artifacts such as floating and ground penetration, and a PD controller that generates joint torques from position and velocity errors. (**C**) Gait-cycle segmentation performed via spectral-peak and extremum detection, illustrated with a representative segmented clip. (**D**) Physics-based correction of foot–ground contact states to eliminate vibration, floating, and penetration artifacts. (**E**) Step-length–speed distribution of the full locomotion dataset, combining captured data from healthy and post-stroke subjects as well as public gait databases. (**F**) Comparison of healthy and patient trajectories in both foot and spinal-top motion, along with visualized walking sequences on the digital human model.

## Data Acquisition and Physics-based Avatar Establishment

We construct a unified gait locomotion dataset and develop a physics-based digital human model to support subsequent analyses. Motion data is captured using five inertial measurement units (IMUs) mounted on the pelvis, thighs, and shanks (Fig. 2A). Signals were transformed from the Earth Coordinate System (ECS) to the Joint Coordinate System (JCS) following the standard formulation described in (24). Detailed participant recruitment and data-collection procedures are provided in the Methods section.

A lower-limb digital avatar is implemented in Unity 3D with PhysX physics simulation engine (Fig. 2B), featuring a six-degree-of-freedom pelvic root and nine actuated lower-limb degrees of freedom controlled via proportional–derivative (PD) torque actuation. Contact constraints, joint coupling, and non-slip conditions were enforced to suppress sensor-induced artifacts such as floating, vibration, and ground penetration. Model parameters and joint specifications are summarized in Table S2 and follow established anatomically consistent musculoskeletal models (25, 26). These parameters were applied uniformly across subjects and tasks, with subject-specific adaptation learned at the control level. The recorded sequences were then denoised and segmented into single-step clips using each subject's dominant gait frequency, followed by physics-based refinement to correct contact artifacts (Fig. 2C–D).

The resulting dataset includes 162 healthy-walking clips and 21 post-stroke clips, totaling 21,674 frames. The joint distribution of step length and walking speed (Fig. 2E) shows that healthy locomotion spans the range exhibited by post-stroke subjects, enabling healthy-calibrated models to adapt to patient-specific gait patterns. At the trajectory level, stroke-related impairments are directly observable in the raw motion data. Representative trajectories (Fig. 2F) highlight impaired-gait characteristics—including reduced step length, asymmetric stance phases, and compensatory circumduction (27, 28). Additional visualizations are provided in Fig. S1.

## Healthy Motion Atlas Performance Evaluation

The control framework integrates reinforcement learning with a physics-based human model to establish a population prior of normative gait, termed the Healthy Motion Atlas (HMA). As shown in Fig. 3A, the controller is a policy network $\pi(a|s)$ optimized using Proximal Policy Optimization (PPO) (29), which maps motion-state and locomotion-goal information to activation signals for the PD controller. The state representation combines locomotion goals, body-state information, and environmental interaction cues (Fig. 3B). The overall simulation–learning workflow links physics-based simulation with neural policy learning (Fig. 3C), while iterative PPO updates progressively refine the controller (Fig. 3D). Detailed policy formulation, state and activation-signal definitions, and training settings are provided in the Methods.

The HMA is trained to capture population-level gait dynamics from healthy-walking clips, encoding coordinated patterns of state ‑ activation evolution. Over 230 million environment steps (≈4 days 8 h), the controller progresses through three characteristic phases: initial standing (0–1 M), transient forward falling (≈1–3 M), and stable stepping emerging beyond 3 M steps. Convergence occurs around 180 M steps, producing a stable, cycle-consistent policy that maintains rhythmic locomotion and serves as a foundation for individualized patient adaptation. The training curves for cumulative reward and sequence length are shown in Fig. 3E.

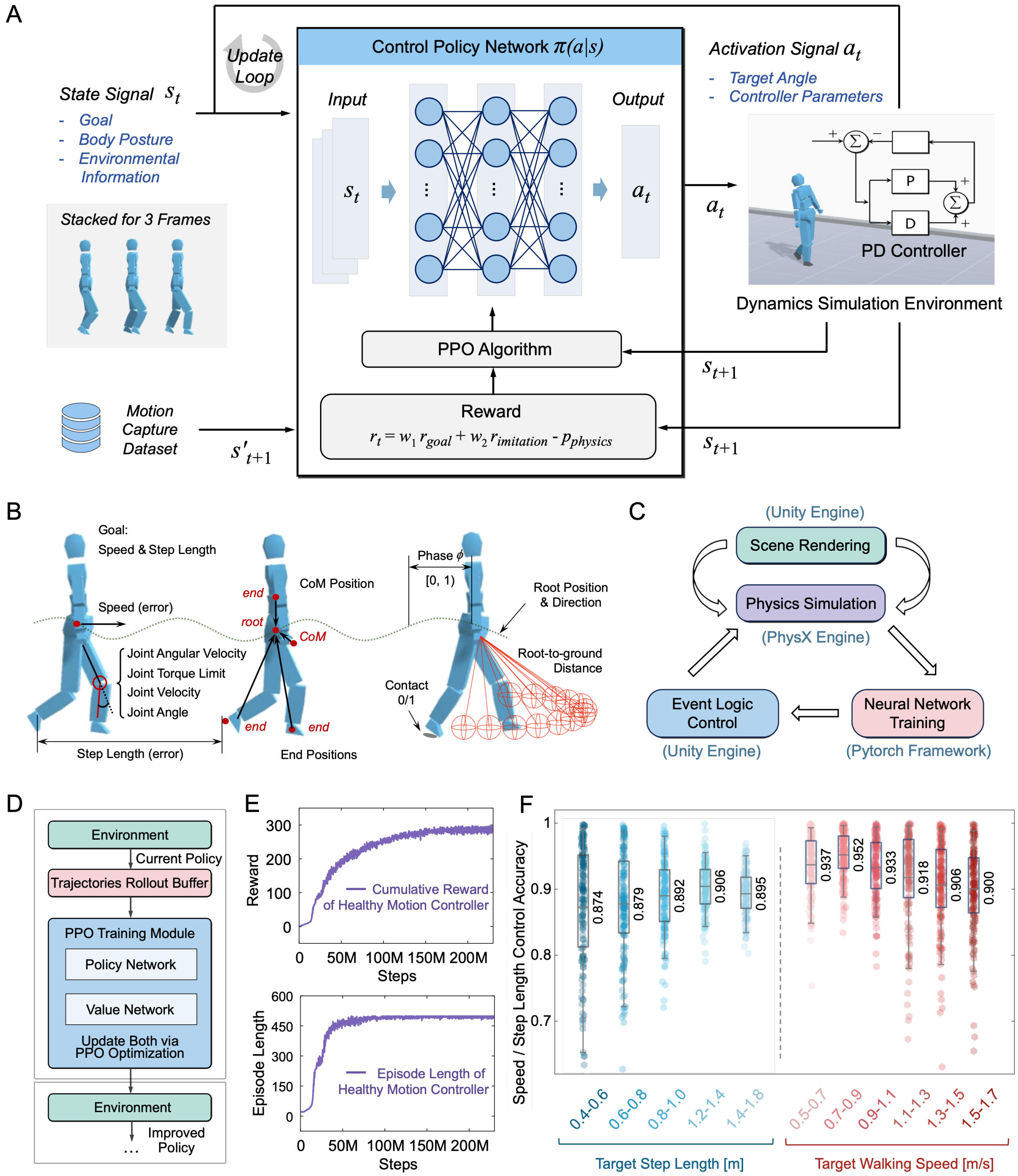

**Fig. 3. Development of the Healthy Motion Atlas model via deep reinforcement learning.** (**A**) Overview of the control-policy module, which receives sequential motion states and outputs proportional–derivative (PD) target signals for a 5-joint model. Training is performed in a closed-loop interaction with the physics environment, where the PPO algorithm updates the policy based on the state transitions and rewards. (**B**) Composition of the state vector, including goal variables (target speed and step length), global gait descriptors, segment kinematics (joint angles, angular velocities, torques), and environment cues such as foot–ground contact and root-to-ground

distance. (**C**) Integrated training pipeline coupling Unity-based scene rendering and PhysX simulation with PyTorch neural-network optimization, where event logic and policy updates are synchronized each frame. (**D**) PPO training loop: the current controlling policy repeatedly interacts with the environment to collect trajectories, which are stored in a rollout buffer and then used by the PPO training module to update both the policy and value networks. (**E**) Training curves showing progressive increases in cumulative reward and episode length as the controller converges. (**F**) Accuracy distribution of the trained Healthy Motion Atlas model across different target step lengths and walking speeds.

To evaluate generalization, we examine performance across combinations of target speed and step length. The distribution of normalized rewards remains broadly uniform within the trained range, with the highest accuracy at moderate speeds (0.5–1.1 m/s) and step lengths (1.2–1.8 m)—patterns consistent with human walking. Sensitivity analyses and control-accuracy maps are presented in Fig. 3F and Fig. S2. Joint- and endpoint-level errors, computed with respect to IMU-based kinematic state estimates, are summarized in Fig. S3, indicating a mean single-axis joint-angle error of 11.61°, a Center-of-Mass (CoM) error of 2.69 cm, and an end-effector error of 12.32 cm. These results demonstrate that the Healthy Motion Atlas achieves stable, biomechanically coherent gait control suitable for downstream clinical adaptation.

### Atlas-to-Patient Clinical Adaptation Performance Evaluation

The Atlas-to-Patient Clinical Adaptation module fine-tunes the HMA to patient-specific gait characteristics through transfer learning (Fig. 4A). The personalized controller inherits the pretrained state–activation policy while adapting to each patient's data, thereby retaining population-level coordination patterns while capturing individual impairments and control dynamics. Atlas-to-patient adaptation and all subsequent personalized evaluations were conducted on the model-assisted patient group ($n = 11$).

Fig. 4B illustrates representative adaptation dynamics and control performance for a single patient, tracking the evolution of multiple reward components, including root position (i.e., avatar's pelvis position) and velocity, the step-length and posture imitation terms, as well as the cumulative reward and overall episode length. Each sub-reward shows gradual stabilization, confirming that the personalized controller progressively aligns with the target walking pattern. Using the trained patient-adapted policy, 500 repeated evaluation trials were conducted under prescribed goals of 1.02 m/s walking speed and 0.80 m step length. The resulting outcomes (Fig. 4C) show mean values of 0.98 m/s and 0.7626 m, with bars representing means and error bars denoting the 25th/75th percentiles. These results indicate accurate tracking of both temporal and spatial gait parameters within the patient's natural rhythm.

Fig. 4D provides a group-level summary across the 11 patients, reporting individual joint- and position-level control accuracy aggregated over repeated walking trials. Mean single-axis joint-angle errors are 5.48° across lower-body articulations, together with 1.67 cm CoM deviation and a 6.35 cm end-effector error. These errors were substantially lower than the corresponding values for the healthy atlas controller (11.61°, 2.69 cm, and 12.32 cm, respectively), indicating that atlas-to-patient adaptation improved patient-specific control accuracy while preserving the coordinated gait prior learned from healthy locomotion. From a rehabilitation perspective, these error measures correspond to different aspects of post-stroke gait control, where joint-angle error reflects the accuracy of lower-limb coordination, CoM deviation relates to postural control and balance regulation, and end-effector error reflects distal limb placement during walking. The reduction of these errors therefore suggests that the adapted avatar better represents patient-specific motor deficits and compensatory movement characteristics, rather than simply reproducing a generic healthy-gait pattern. This individualized representation provides the basis for the subsequent task-conditioned prediction stage, because slope ascent and stair climbing require the model to preserve patient-specific gait characteristics while adapting them to new terrain demands. Full learning-curve, performance, and efficiency comparisons between atlas-initialized and non-initialized training are provided in Table S3 and Fig. S4, demonstrating that atlas initialization substantially accelerates convergence while maintaining biomechanical coherence in activation signal patterns and kinematic gait characteristics.

## Terrain-Adaptive Locomotion Prediction

The locomotion-prediction framework combines behaviour cloning (BC) and generative adversarial imitation learning (GAIL) (30) with physics-based reinforcement learning (Fig. 5A). BC provides early stabilization of state–activation alignment, while GAIL regularizes the generated control signals to maintain both physiological plausibility and patient-specific rhythm across tasks. We evaluate the framework on clinically representative rehabilitation tasks of increasing difficulty—slope ascent and stair climbing—using a curriculum that progressively increases task demands. Illustrative results are presented for a representative ambulatory stroke patient with left-sided impairment and preserved independent mobility, with nominal level-ground parameters of 0.707 m/s walking speed and 0.801 m step length.

*Task I - Slope Ascent*

Training on the slope-ascent task required approximately 23 million environment steps ($\approx$ 8 h). Across 5 curriculum levels, both task difficulty and associated reward components (Fig. 5B and Fig. S5A) showed clear progression and stable convergence. Importantly, the controller consistently preserved the patient's characteristic gait features, including limited knee flexion, circumduction during swing, and compensatory pelvic hike, despite increasing terrain demands. The training steps required to complete successive curriculum levels increased with difficulty (0.7 M, 2.3 M, 4.7 M, 5.9 M, and 9.4 M), reflecting the greater control complexity of steeper slopes.

GAIL-related training diagnostics, including the internal imitation reward, discriminator output, and discriminator loss, further indicated stable optimization (Fig. S6). The discriminator output remained close to 0.5 throughout training, indicating close alignment between generated activation patterns and the patient's flat-ground activation patterns as the controller adapted to the slope task. To quantify this alignment during training, we use an activation-signal imitation similarity score, which measures pattern-level similarity between generated task-conditioned and patient flat-ground activation sequences; higher values indicate closer alignment. Under slope conditions, this score reached 99.70% (details in Table S4), showing that the generated slope-ascent activation patterns remained highly similar to the patient's flat-ground pattern. Note that this score is not based on real-world task-specific ground truth; rather, it is used internally during training to characterize cross-task activation consistency.

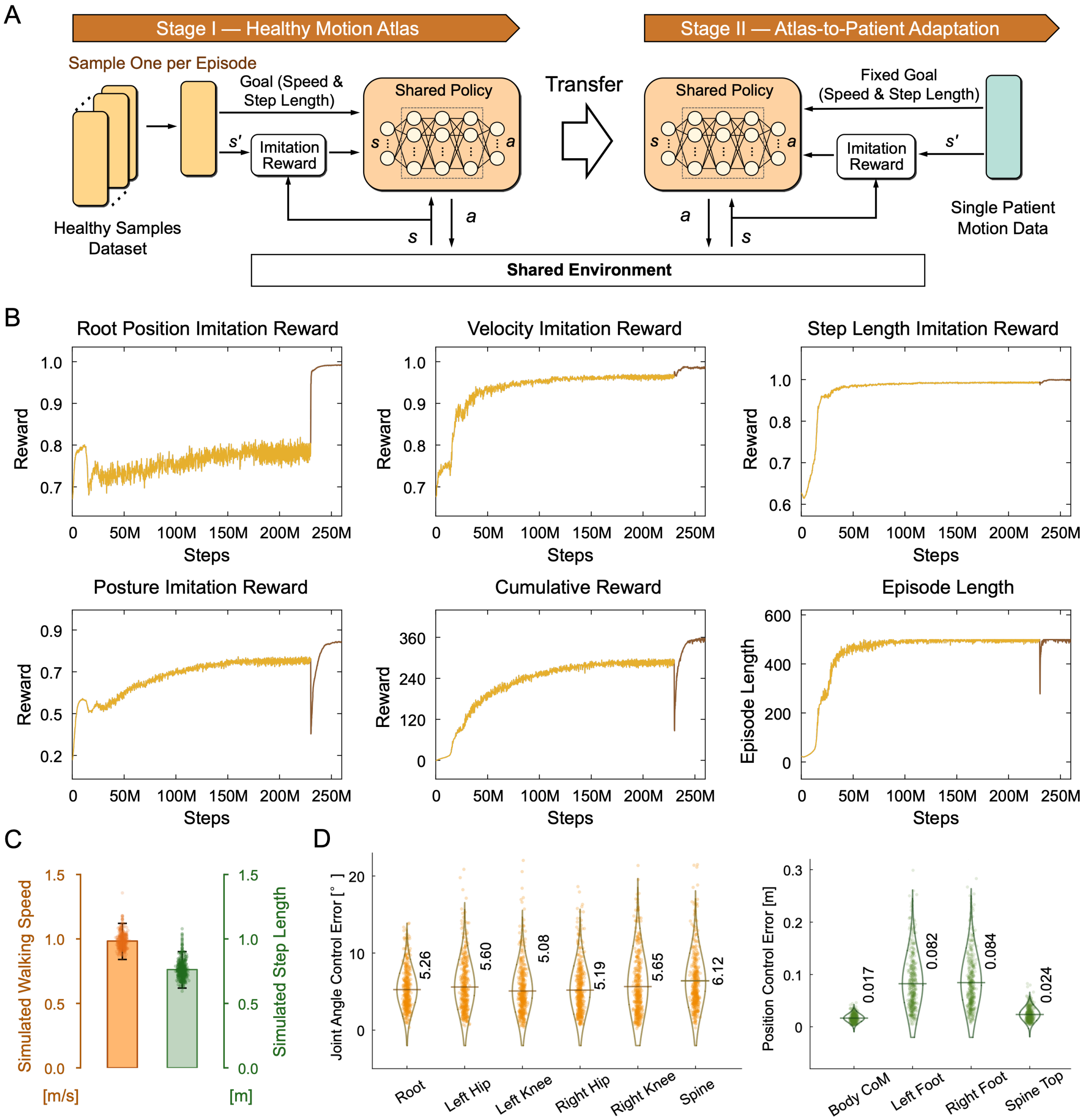

**Fig. 4. Atlas-to-patient adaptation via goal-conditioned transfer learning**. (**A**) Adaptation scheme in which the Healthy Motion Atlas initializes the controller, and personalization is achieved by updating task goals (target speed and step length). (**B**) Training curves showing rapid gains across all imitation objectives and cumulative reward, while episode length remains consistently high, indicating fast personalization and stable control performance enabled by the Healthy Motion Atlas. (**C**) Comparison between the representative patient's measured gait parameters and the personalized controller's outputs, showing high-fidelity reproduction of walking speed (1.02 m/s vs. 0.98 m/s) and step length (0.80 m vs. 0.76 m). (**D**) Distributions of joint-angle control error across all lower-body articulations, together with position-error distributions for the body center of mass and distal end points under the personalized controller summarized across the model-assisted group.

*Task II - Stair Climbing*

For stair climbing, we adapt a five-level curriculum with progressively increasing step height. Under curriculum progression, the cumulative reward and episode length grow smoothly and then plateau, indicating stable mastery (Fig. 5B). Step budgets also increase with difficulty (11.3 M, 13.4 M, 26.1 M, 43.9 M, and 57.3 M). Detailed training curves of all reward components are provided in Fig. S5B. To address the known difficulty of producing natural stair gaits in example-guided DRL (31), we introduce a minimal terrain-aware guidance reward that encourages a leg-raising action when a step is detected ahead; we also report a +guidance/–guidance ablation. Without this cue, the policy fails to mount the stairs and oscillates near the edge, underscoring the need for task-aware shaping. The definition of this guidance term and additional analyses are provided in Methods. Fig. 5C shows the intrinsic reward and discriminator output during GAIL-based stair adaptation. Using the same activation-signal imitation metric, the stair-task consistency reaches 79.72% (Table S5).

The data-physics hybrid controller shows strong robustness to external perturbations. Perturbation tests (31, 32) (Table S6) indicate that it withstands forward disturbances up to 980 N and lateral disturbances up to 320–360 N while maintaining balance. A baseline comparison (Fig. S7; Table S7) shows that the proposed data–physics hybrid converges on a 1:12 slope in approximately 23 million training steps, whereas a DeepMimic-style reference-tracking controller (31), trained to imitate a single posture sequence under physical constraints, requires approximately 90 million steps to achieve comparable stability. This threefold improvement in training efficiency underscores the effectiveness of the proposed unified control framework. As a result, the proposed method reduces training time to 25.56% of the baseline while achieving +4.73% higher joint-angle imitation accuracy and +12.10% higher distal-endpoint accuracy. The observed performance suggests that physiologically consistent generalization benefits from modeling underlying control activation patterns within a generative control framework, rather than relying on motion imitation alone.

*Prediction Accuracy Across Patients and Tasks*

Fig. 5D reports average patient-specific task posture prediction accuracy across the model-assisted group (n = 11), by comparing generated motions with measured kinematics in real-world tasks. Prediction accuracy was quantified using the posture similarity metric defined in the Methods, based on joint-angle deviations aggregated across all articulated joints over the motion sequence; higher values indicate closer agreement between generated and measured task motions. For each difficulty level, similarity scores were summarized across patients as mean ± s.e.m. Statistical significance relative to reference baselines was assessed using two-sided paired *t*-tests. Significance levels are indicated as $p < 0.05$ (*) and $p < 0.01$ (**). Two reference levels are shown to contextualize the prediction accuracy. First, an 87% simulation reconstruction accuracy is shown as an upper reference, obtained by reconstructing recorded patient motion in physics-based simulation. This reflects unavoidable wearable-sensor noise, drift, and minor variations in sensor placement and body dimensions, together setting a practical upper bound for posture comparison. Second, a flat-to-task gait deviation reference is shown, computed using the same posture similarity metric, to quantify the natural gait change when transitioning from flat-ground walking to slope or stair locomotion (averaging ~75% for slopes and ~48% for stairs).

Across the 11 patients, the results show 82.2% ± 1.6% posture similarity on slopes and 69.9% ± 2.1% on stairs. These values correspond to approximately 95% and 80% of the maximum attainable accuracy under noise- and model-limited conditions, demonstrating accurate prediction of task-specific gait patterns well beyond the natural deviation from flat-ground walking and close to the simulation reconstruction limit. Supplementary Table S8 further reports joint-profile Pearson correlation coefficient (PCC), which measures waveform-level agreement between generated and measured joint-angle trajectories after time normalization; values closer to 1 indicate stronger correspondence. Joint profile PCC ranged from 0.90 ± 0.05 to 0.93 ± 0.04 for slope ascent and from 0.87 ± 0.06 to 0.91 ± 0.04 for stair climbing across difficulty levels. PCC values above 0.70 are commonly interpreted as indicating strong correlation in medical research (33). The observed PCC values therefore indicate strong waveform-level agreement between predicted and recorded joint profiles.

Fig. 5E shows representative visualized rollouts across difficulty levels (simulated task completion vs. simulated task failure) for slopes and stairs, illustrating how the system visualizes success and indicates simulated motor challenge; for example, low stair height (7.5 cm) yields stable ascent, whereas high stair height (15 cm) induces excessive lateral trunk sway (56.1°), signaling a simulated task-performance boundary for the current stage. Building on the individualized locomotor representation learned during atlas-to-patient adaptation, the locomotion predictions characterize how each patient-specific avatar responds to terrain-dependent task conditions beyond the observed level-ground walking scenario. Conventional assessment describes current gait performance, but does not directly reveal how a patient may adapt to increased functional mobility demands, such as incline walking or stair climbing. By simulating these task conditions with the patient-adapted avatar, the framework extends the observed level-ground assessment to unmeasured rehabilitation-relevant scenarios, providing scenario-specific information on task feasibility, balance-related movement adaptation, and compensatory gait strategies. These simulations indicate how a patient may respond to increasing task demands—whether task performance is likely to be maintained and what compensatory adaptations may emerge. Given that these projections are simulation-based rather than directly measured, they are intended as decision-support information rather than direct clinical risk estimates or automated prescriptions.

## Clinical Integration and Rehabilitation Outcomes

To evaluate clinical usability, the proposed data–physics hybrid generative model is deployed across the five rehabilitation institutions (Fig. 6A). At each site, the workflow followed a unified clinician-in-the-loop protocol. In the model-assisted group, the 20 m walking trial was repeated daily to refresh the patient-specific simulations for ongoing rehabilitation planning. Using each updated clip, the system generated patient-specific simulations for slope ascent and stair climbing tasks. Therapists at each hospital reviewed the predicted gait trajectories and the associated activation-signal visualizations to examine whether the patient could maintain stability, where compensatory movements might emerge, and how task difficulty should be adjusted. The system did not prescribe exercises; instead, it served strictly as a decision-support tool, allowing clinicians to incorporate model-generated insights into their own professional judgment. This ensured that therapist expertise—patient fatigue, comorbidities, motivation, and safety considerations—remained central to individualized treatment planning across all five sites. Therapist-reported feedback further indicated that the system integrated well into routine clinical workflows, imposed acceptable time burden, and provided clinically interpretable and actionable information for rehabilitation planning (Fig. 6C).

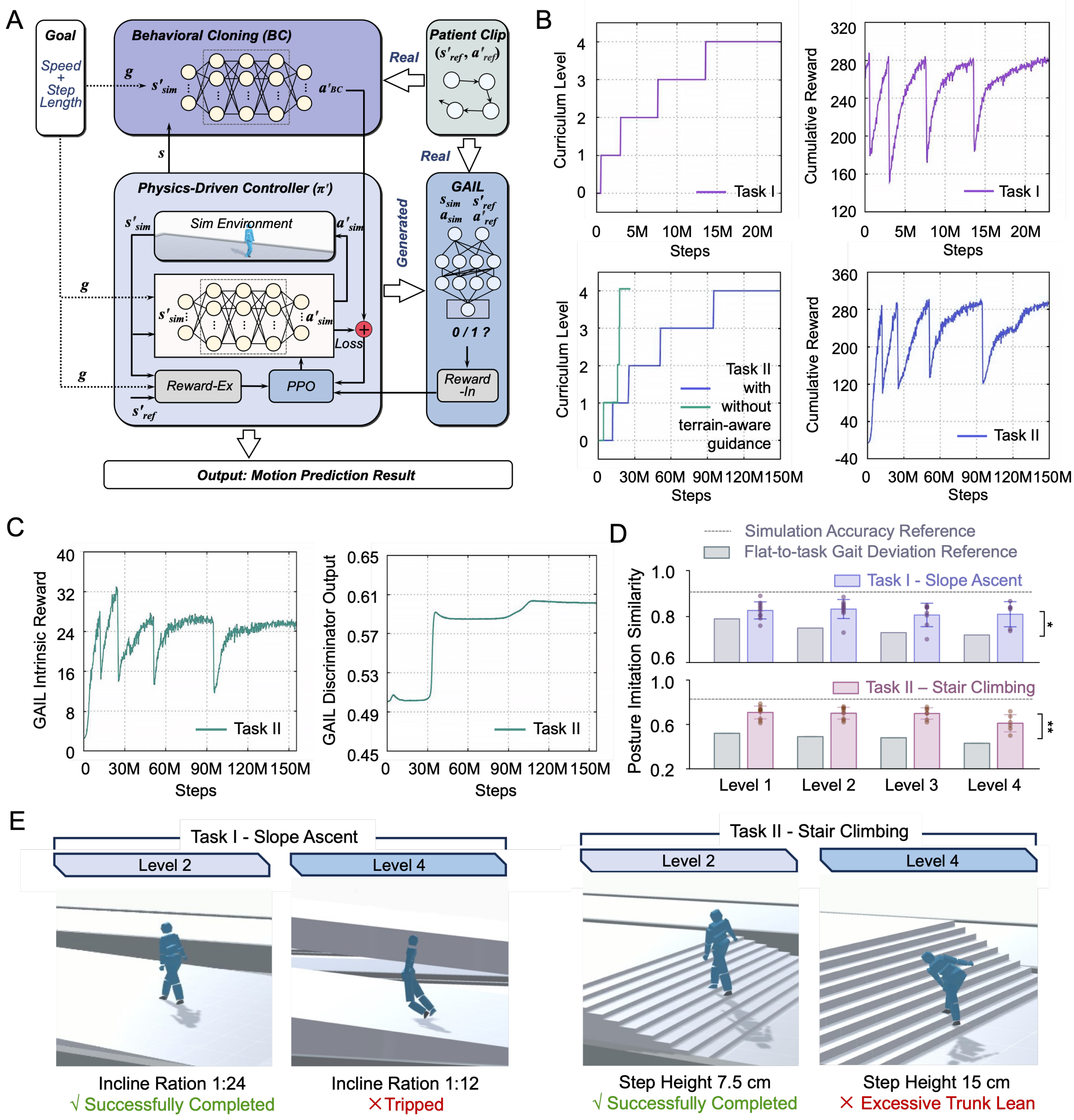

**Fig. 5. Terrain-adaptive locomotion prediction and rehabilitation plan design.** (**A**) Data–physics hybrid generation framework that operates on patient-specific data, combining a personalized physics-driven controller with behavioral cloning and a GAIL network to maintain consistent activation patterns across motion; curriculum tasks are defined across difficulty levels spanning slopes and stairs. (**B**) Curriculum progression and reward curves during training for slope and stair tasks, indicating the relative difficulty and learning time required at each level. (**C**) GAIL intrinsic reward and discriminator output during stair-task adaptation, showing improved alignment between generated and real motion patterns, with imitation consistency reaching 79.7%. (**D**) Posture prediction accuracy evaluation across 11 personalized models, comparing predicted motions with recorded patient trajectories under different tasks and difficulty levels. Significance

is evaluated relative to the task-dependent lower-bound baseline, with $p < 0.05$ (*) and $p < 0.01$ (**); corresponding values are $p = 0.032$ for Task I and $p = 0.004$ for Task II. (**E**) Visualized rollouts illustrating example outcomes across difficulty levels, showing stable execution at appropriate difficulty and the emergence of compensatory or unstable behaviors (such as tripping or excessive trunk lean), when the task becomes too demanding.

A randomized pilot feasibility study was then conducted to examine the clinical integration of this deployment. The same 21 post-stroke patients with lower-limb motor impairment (model-assisted, n = 11; control, n = 10) received lower-limb rehabilitation for 28 days, with weekly FMA-lower-extremity (FMA-LE) assessments (34) performed by blinded therapists (Fig. 6B). The rehabilitation program provided approximately three hours of supervised therapy each day, comprising balance and postural-control training, lower-limb strengthening, range-of-motion and stretching exercises, and walking practice. Among these components, walking practice accounted for approximately 50% of total daily therapy time and included overground walking, slope ascent, and stair climbing. Patients were randomized within each site, and treating therapists at each site provided care for participants in both study groups, limiting site- and therapist-level imbalance. Overall rehabilitation exposure was comparable between groups, indicating that the model-assisted group did not receive a systematically higher rehabilitation dose. This included similar total therapist contact time ($p = 0.91$) and total walking practice (50.3 ± 2.1% vs. 49.7 ± 1.8%, $p = 0.49$). Within walking practice, clinicians in both groups selected tasks and difficulty before each daily session according to each patient's current motor ability, including whether to continue with overground walking alone or introduce slope ascent and stair climbing, and at what difficulty level.

In the model-assisted group, clinicians additionally incorporated patient-specific avatar simulations into this decision-making process (Fig. 6A). Detailed session-level rehabilitation exposure and between-group comparisons are provided in Supplementary Table S9. In the model-assisted group, the same workflow was repeated daily using updated 20 m walking assessments to refresh patient-specific simulations for ongoing rehabilitation planning. At baseline, impairment levels were comparable between groups (control: 26.0 ± 0.4; model-assisted: 25.6 ± 0.6). Over the four-week period, however, patients whose therapists used the proposed generative model's predictions exhibited faster functional improvement. By day 28, the model-assisted group achieved an average FMA-LE score of 31.6 ± 0.6, whereas the control group reached 29.7 ± 0.5. Between-group differences in FMA-LE scores were evaluated using a two-sided independent-samples t-test and were statistically significant ($p < 0.05$). Consistent with these findings, patients reported high acceptability of the wearable-based 20 m walking assessment, with favorable ratings for comfort, safety, and willingness to repeat the procedure during rehabilitation (Fig. S8).

These results demonstrate that the proposed dynamic, task-predictive generative model provides clinicians with actionable, scenario-specific insights that conventional static assessment methods cannot offer. By enabling therapists to preview patient-specific performance in slope and stair tasks, the system supports individualized task planning, and helps better align task difficulty with each patient's motor capability. This clinician-in-the-loop decision-support process may help shape more individualized and efficient recovery trajectories in lower-limb post-stroke rehabilitation.

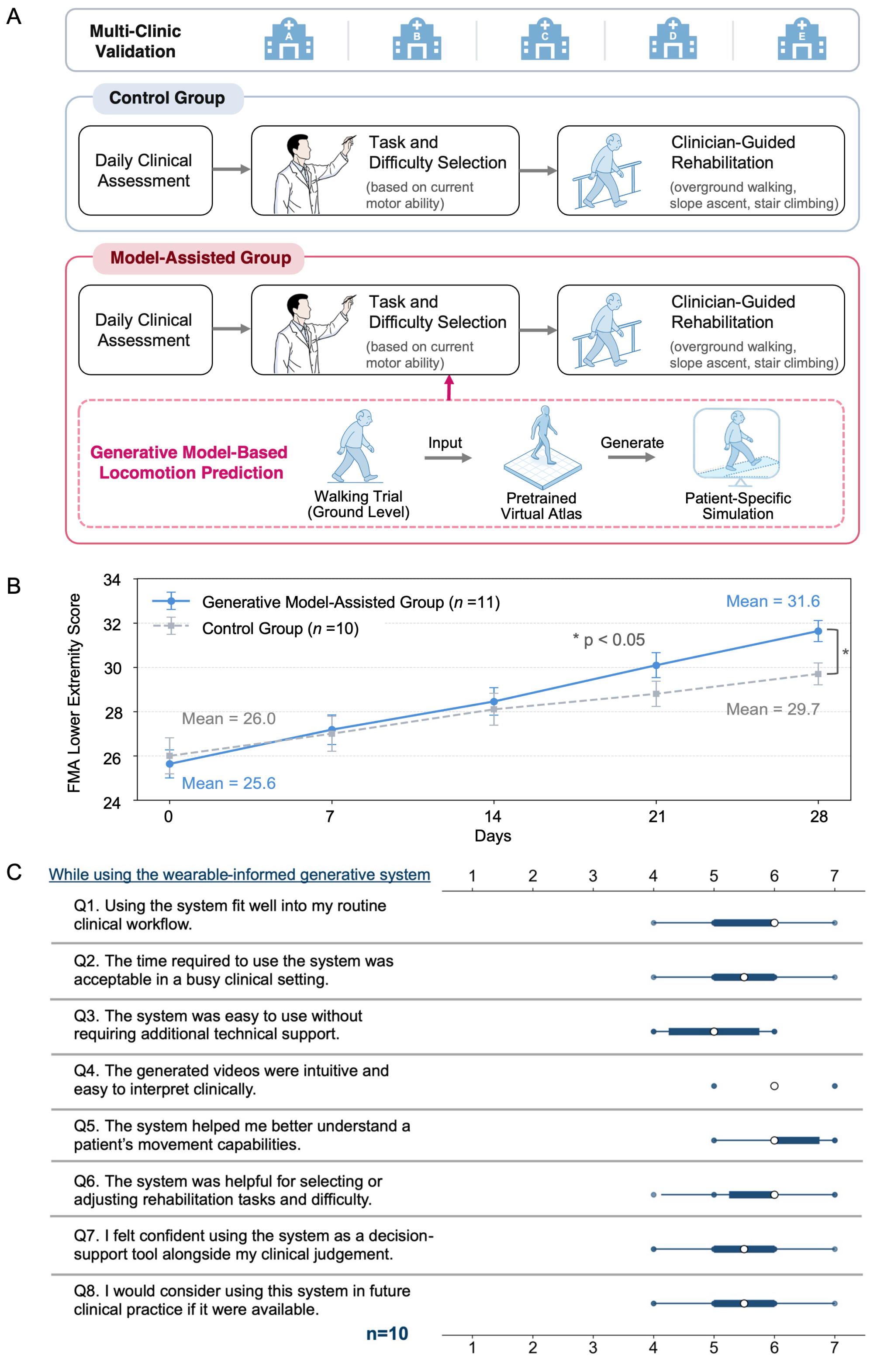

**Fig. 6. Clinical integration and rehabilitation outcomes.** (**A**) Clinical rehabilitation workflow for both groups, highlighting deployment of the data–physics hybrid generative model in the model-assisted group. At the start of each daily rehabilitation session, gait data are collected during a 20 m walking test to construct or refresh the individualized model. The system simulates rehabilitation tasks of varying difficulty to provide additional patient-specific information for clinician-in-the-loop treatment planning. (**B**) Pilot randomized study in 21 inpatients with post-stroke lower-limb impairment (generative-model-assisted group, $n = 11$; control group, $n = 10$). All patients underwent 28 days of lower-limb rehabilitation, with task and difficulty progression individualized according to each patient's motor ability in both groups; only therapists in the model-assisted group had access to model predictions during planning. Fugl-Meyer Assessment lower-extremity (FMA-LE) scores (mean ± s.e.m.) improved in both groups, with larger gains in the model-assisted group (baseline: 25.6 ± 0.6 vs 26.0 ± 0.4; day-28: 31.6 ± 0.6 vs 29.7 ± 0.5; $p = 0.029$ between groups at indicated time points), suggesting that generative-model-based decision support may enhance functional recovery. (**C**) Therapist feedback collected after using the system to support rehabilitation planning, rated on a 7-point Likert scale (1 = strongly disagree; 7 = strongly agree) across eight items covering workflow integration, time burden, ease of use, interpretability of generated videos, perceived clinical usefulness, decision-support confidence, and adoption intention. Dots represent individual responses; thick bars denote the interquartile range (IQR), whiskers indicate 1.5 × IQR, and white circles mark the median. Higher scores indicate greater usability and perceived clinical value. n = 10 therapists.

## III. Discussion

This study advances a new paradigm for wearable-based motor rehabilitation by shifting wearable data processing from static, descriptive assessment to dynamic, visualized task performance prediction. Rather than treating wearable gait signals as endpoints for static scoring or descriptive analytics, we demonstrate how a brief, wearable-captured walking sequence can be transformed into a patient-specific, generative representation of locomotor capacity across unobserved rehabilitation tasks. By integrating wearable-derived kinematics with a physics-based control model, a population-level Healthy Motion Atlas, and goal-conditioned reinforcement learning, the proposed framework reconstructs individualized locomotor control signatures and uses them to generate dynamic predictions for clinically meaningful mobility challenges such as slope ascent and stair climbing. In doing so, wearable data are elevated from passive measurements that underpin static, metric-based predictions to active drivers of dynamic, task-conditioned decision support grounded in physically plausible motion. The resulting simulations preserve patient-specific movement style, achieve high fidelity at joint and endpoint levels, and remain physically consistent, while requiring only minimal wearable input. Together with multi-center pilot results showing significantly improved Fugl–Meyer lower-extremity outcomes when these predictions are used in planning, this work establishes a practical and scalable wearable-centered framework for anticipating task feasibility, progression difficulty, and personalized intervention in post-stroke mobility rehabilitation.

These pilot clinical findings were obtained within a feasibility-study design. FMA-LE outcomes were assessed by therapists blinded to group allocation. Treatment prescription could not be fully blinded, because the model-assisted intervention necessarily required therapists to review patient-specific simulation results. To limit the potential confounding influence of this non-blinded treatment step, randomization was performed within each clinical site, and both groups received the same overall rehabilitation dose throughout the treatment period, with consistent proportions across the major exercise categories. Within this controlled treatment framework, the proposed model was intended to support patient-specific adjustment of task difficulty, rather than to systematically increase rehabilitation intensity or dose. This rationale is consistent with the motor-learning principle that task difficulty should be matched to a patient's current motor ability and tolerance to support effective practice (35, 36). Taken together, these findings support the feasibility of using generative model-assisted individualized rehabilitation planning within rehabilitation care, while larger prospective trials will be needed to further evaluate its generalizability and refine treatment-delivery protocols.

At the same time, it is important to situate these results within the broader context of gait modeling and prediction. Clinically reliable prediction of human locomotion has long been recognized as challenging (37), even when using substantially richer sensing modalities than wearable IMUs, such as marker-based motion capture, force plates, and electromyography (25, 38). Prior neuromusculoskeletal modeling studies have reported persistent limitations related to inter-subject variability, sensitivity to subject-specific calibration, and accumulated estimation errors (38, 39), which together constrain robustness and clinical transferability. In this context, the present work does not seek to replace high-fidelity biomechanical models, but rather explores a complementary, wearable-centered paradigm that enables stable and interpretable predictions under sparse sensing.

Several considerations temper the interpretation of these findings. First, our evaluation focused exclusively on two rehabilitation scenarios, slope walking and stair climbing. These tasks are among the most common and clinically consequential in gait retraining (40, 41), yet they do not encompass the broader range of activities encountered in rehabilitation practice, such as turning, dual-task walking, robot-assisted gait training, or perturbation-based balance tasks (42). Extending the model to these scenarios will require expanding the controller's action space and incorporating terrain- or device-specific dynamics. Second, our current patient cohort consists of stroke survivors who are able to walk independently without assistive devices. The present framework is therefore tailored to this functional subgroup and does not directly generalize to individuals who rely on canes, walkers, or caregiver support. Extending the system to these populations will necessitate explicit modeling of assistive-device interactions, more complex human – device dynamics, and larger multi-center datasets to capture broader clinical variability. Third, as with any simulation-based prediction, performance in novel tasks cannot be directly verified at the moment of use. The system should therefore be understood as a decision-support tool rather than a replacement for clinical judgment, and its task recommendations should be interpreted in conjunction with therapist expertise and patient-specific constraints. Future prospective trials will be required to calibrate uncertainty, validate simulation-derived task thresholds, and establish evidence-based standards for clinical adoption.

Despite these limitations, the proposed framework holds significant promise for the future of neurorehabilitation. By enabling low-cost, cross-clinic prediction of task-specific motor capability from minimal data, the method may improve personalized treatment planning for stroke and other neuromuscular or neurodegenerative disorders. Because the controller embeds musculoskeletal and neuromotor constraints within a physics-based environment, the underlying principles can, in concept, be adapted to conditions such as Parkinson's disease, cerebral palsy, and spinocerebellar ataxia with appropriate task and data adjustments. In this sense, the approach can be viewed as an early step toward a human body digital twin for lower-limb motor function, in which continuously updated, individualized simulations mirror a patient's evolving gait capacity and constraints (43). Moreover, as the system is deployed in practice, its continuous use could be integrated with active-learning strategies to iteratively refine the underlying models, progressively building a richer, rehabilitation-centered representation of human motor behavior. Such a data-driven evolution could generate new insights into disease mechanisms, task difficulty progression, and recovery trajectories. In the longer term, once robustness and generalization are established, this modeling paradigm may support scalable, remotely delivered rehabilitation guidance in under-resourced or underserved regions, helping to democratize access to personalized, high-quality neurorehabilitation worldwide.

# IV. Methods

## Data Collection Protocol and Participant Recruitment

To support individualized controller learning and task-conditioned locomotion prediction, we constructed a lower-limb gait dataset from controlled clinical trials and a supplementary public motion-capture repository. Participants for the clinical component were recruited from Beijing Tsinghua Changgung Hospital, Beijing Tongren Hospital, Shijiazhuang People's Hospital, Beijing University of Chinese Medicine, and the Fifth Medical Center of the Chinese PLA General Hospital (301 Hospital). The patient cohort comprised 21 individuals with post-stroke hemiparesis in the mid-recovery phase (1–6 months post-stroke), all of whom were able to stand and walk independently. Demographic variables—including age, sex, height, weight, affected side, and time since stroke—are reported in Table S10. To provide reference locomotion patterns, an additional 19 healthy adults were recruited and assessed using the same data-collection protocol.

All clinical participants completed habitual-pace straight walking along a 20 m level walkway. Lower-limb kinematics were recorded using five Perception Neuron 3 nine-axis IMUs placed on the pelvis, bilateral thighs, and bilateral shanks (90 Hz sampling; 0.02° pose resolution) (44). Sensor placement was standardized across individuals to ensure consistent measurement of segment rotations and translations. To broaden dataset coverage and to enrich the representation of high-quality nominal gait cycles, we further incorporated 67 straight-walk sequences from the CMU Motion Capture database (optical capture at 120 Hz) (45). Statistics for each cohort—including the number of clips, average duration, sampling rates, and cycle-to-cycle consistency—are summarized in Table S11. Following the acquisition, all recordings were denoised and segmented into single-step clips using the processing pipeline described in the Results Section. The clinical study received ethical approval (project no. 2020036), and the study involving healthy participants was approved by the Beihang University Ethics Committee. Written informed consent was obtained from all participants. No participants were lost to follow-up or excluded from the clinical analysis.

## Control Policy Formulation

We define the avatar's control policy $\pi\ (a_t \mid s_t)$ that outputs the motor activation $a_t$ given the simulated state $s_t$ at time $t$. The policy is optimized to generate motions that match the reference posture $P_t$ at each time step. The human motor-control system is highly complex and underdetermined, making it impractical to derive a closed-form equation for the activation signal $a_t$. To address this, we employ DRL techniques to learn a control policy network (46) through iterative trial-and-error interaction with the environment, during which the policy updates its parameters based on feedback from each $s_t - a_t$ transition.

Our model builds upon the Deepmimic framework proposed by Peng et al. (31), a widely used framework in physics-driven motion generation tasks due to its robustness and high fidelity in mimicking reference actions. In this study, we expanded its inputs beyond fixed body postures to include arbitrary goals like walking speed, step length, and cadence, as depicted in Fig. 3A. This extension enables the model to learn diverse walking patterns from healthy individuals, thereby accelerating the adaptation process and facilitating personalized control-policy training for patients.

*States and Activation Signals*

The input state signal $s$ comprises five main components, as illustrated in Fig. 3B: (i) phase [0, 1), which aligns the current frame to the target frame for sequence imitation and removes the need to feed target actions directly, improving compactness and scalability (47); (ii) motion goal (target walking speed, step length, and their deviations from the current state); (iii) PD control information (per-joint torque upper limits); (iv) body posture: root height/rotation, each child node's local rotations plus linear/angular velocities, as well as CoM and limb-endpoint positions relative to the root (all angles in quaternions); and (v) environment interaction: foot–ground contact (0/1), hip–ground distance, and terrain rays cast near the root over the past three frames. Rays span the sagittal plane from –90° to 90°, cast at 45° increments (13 rays total). This range-sensing approach is more efficient than height maps or image inputs while still supporting complex terrain (e.g., stairs) and improving stability. In total, the state vector comprises 122 motion state variables.

The activation signal $a$ consists of two parts: target rotation angles for each joint in the PD controller, and the upper limits of joint torques. The dimensions of the target rotation angles align with the degrees of freedom (DoF) of the controllers, while the joint torque limits specify the maximum torque for each joint, totaling 14 dimensions.

*Reward*

During policy training, each action $a_t$ produced by the controller is evaluated through a reward signal, which provides feedback on how desirable the action is for the task and guides the improvement of the policy over time. The reward function $r_t$ follows prior work (31, 48), and is formulated as a weighted sum of a goal-completion reward $r_{goal}$, a locomotion-imitation reward $r_{imitation}$, and a physical-feasibility penalty $p_{physics}$. However, instead of computing the goal reward $r_{goal}$ independently at each frame as in prior work, we evaluate each goal parameter over a short temporal window. This aggregation over time reduces frame-to-frame fluctuations and mitigates cumulative drift in the controller. The reward function $r_t$ is defined as:

$$r_t = w_1\, r_{goal} + w_2\, r_{imitation} - p_{physics}$$

where $w_1$ and $w_2$ are both set to 0.5. In this formulation, $r_{goal}$ matches global gait targets (walking speed, step length, forward direction, and root pose stability), $r_{imitation}$ preserves local style and coordination (joint angles/velocities, CoM, and distal endpoints), and $p_{physics}$ penalizes non-physical contacts (e.g., invalid foot sliding and mid-air floating) with early termination on non-foot ground contacts. Full mathematical definitions are provided in Supplementary Note S1.

## Atlas-to-Patient Control Model: Policy Representation and Optimization

*Policy Representation*

The control policy $\pi$ is represented by a feedforward neural network that maps a given state $s$ to a motor-activation signal $a$. The signal $a$ is modeled as a Gaussian distribution $a \sim N(\mu(s), \Sigma)$, where the mean $\mu(s)$ is produced by the network, and the covariance matrix $\Sigma$ is assumed diagonal and parameterized as $\Sigma = \varphi I$, where $I$ is the identity matrix and $\varphi$ is a hyperparameter controlling action noise. The network contains three hidden layers with 256 units each, using ReLU activations throughout. The policy input is a 122-dimensional state vector; to enrich temporal observation, we stack three consecutive frames ($t$-2, $t$-1, $t$), yielding a 366-dimensional final input. The policy outputs a 14-dimensional motor activation signal vector.

During network training, a current state $s_t$ is fed into the policy network $\pi$ to obtain the activation signal $a_t$, which is then applied to the joint PD controllers for forward dynamics simulation, producing the next state $s_{t+1}$. The state $s_{t+1}$ is compared with the reference data to compute the reward $r_t$ at time $t$. The training objective is to learn a policy that maximizes the expected cumulative reward over each rollout. To optimize the policy $\pi$, we employ the Proximal Policy Optimization (PPO) algorithm (29), a widely used and robust method for continuous-control problems. This process involves training both the policy network and a separate value function network. The value network is trained in parallel to estimate state values; it is also a feedforward neural network with two hidden layers of 128 units each, using the same input state representation but producing a single scalar value estimate.

*Training Procedure*

Network training proceeds in two stages: Healthy Motion Atlas Construction (HMA) and Atlas-to-Patient Clinical Adaptation (A2P-CA). Simulation and policy execution run at 30 Hz. Each episode is capped at 500 time steps and terminates early upon reaching the cap or detecting a fall. PPO updates occur after collecting 204,800 samples, using a batch size of 2,048 and a reward discount factor of 0.995. All remaining settings (e.g., update cadence, learning-rate schedules) are listed in Table S12.

- *Stage I — Healthy Motion Atlas Construction*

We first construct a normative motion atlas from 162 healthy walking clips. At the beginning of each episode, a target walking speed and step length are randomly sampled from training ranges (0.4–1.4 m/s for speed and 0.5–1.7 m for step length). A reference clip is then selected from the healthy gait dataset as the one whose (speed, step length) pair is closest to the sampled target in Euclidean distance. The reference clip is then time-resampled so that its walking speed and total number of frames match the sampled target. This stage yields a healthy baseline model capable of generating gait patterns across a broad range of locomotion conditions.

- *Stage II — Atlas-to-Patient Clinical Adaptation*

The atlas policy is then adapted to each stroke patient via transfer learning, reducing time and computation while improving generalization, compared with training a patient-specific model from scratch. In this stage, the imitation reference is replaced with a patient-specific walking clip, so the policy learns to match the individual's gait patterns rather than the healthy atlas. Accordingly, the target walking speed and step length in the state $s_t$ are fixed to the patient's observed values. This adaptation preserves the atlas-level motion structure while enabling the controller to capture patient-specific characteristics essential for personalized rehabilitation.

**Personalized Locomotion Prediction Framework**

*Data-Physics Hybrid-Driven Model Architecture*

Building on the individualized control policy, we propose a data-physics hybrid-driven motion generation framework for a range of rehabilitation tasks. Existing DRL-based motion frameworks often rely on hand-crafted rewards and heavy tuning, and manual interventions that may introduce bias. To mitigate these issues, we combine behavioural cloning (BC) (49) for warm starts with GAIL for continual consistency checks of the patterns between generated and real activation signals. Additional context on the broader motion-generation landscape relevant to this framework is provided in Supplementary Note S2.

For different rehabilitation tasks, we make only minor adjustments to the model's states, activation signals, and rewards:

- *State $s'$:* Add or modify only task-specific goal information (e.g., distance-to-path for trajectory following, or adjusted target speed).

- *Activation signal $a'$:* Retain the same definition as in the individualized controller.

- *Reward $r'$:* Keep the original structure; only update the global targets in $r_{goal}$ (such as speed, step length, direction) according to the requirements of each task. For challenging tasks, sparse auxiliary cues may be introduced during early training to aid convergence, with negligible effect on final behavior.

This design minimizes manual engineering, improves physiological interpretability, and enhances predictive flexibility in complex and unfamiliar scenarios.

*Training Procedure*

Locomotion prediction is learned from a single motion clip of each patient. For a given rehabilitation task $g$, the original state and activation sequences $\{s, a\}$ are transformed into their task-conditioned forms $\{s', a'\}$. For clarity of notation in this section, we denote the patient's transformed reference data as $\{s'_{ref}, a'_{ref}\}$. During model training, simulated rollouts generated by the policy are written as $\{s'_{sim}, a'_{sim}\}$, and the learnable policy network is denoted by $\pi'$. The training procedure is divided into two phases: behavioral prior distillation and generative adversarial task prediction.

*- Stage I — Behavioral Prior Distillation (BPD)*

In BPD, a behavior-cloning proxy $\pi_{BC}$ is trained in a supervised manner on the patient's reference clip, using $s'_{ref}$ as inputs and $a'_{ref}$ as outputs. The trained proxy is then duplicated to initialize the policy network $\pi'$, including architecture and parameters, thereby injecting a patient-specific behavioral prior before locomotion prediction. When the rehabilitation setting satisfies $s'=s$ (i.e., the policy's observation space already matches the reference), this phase can be skipped by directly setting $\pi'=\pi$.

*- Stage II — Generative Adversarial Task Prediction (GATP)*

GATP adapts the initialized policy $\pi'$ to the rehabilitation task through a combination of behavioral imitation, GAIL-based adversarial learning, and physics-driven reinforcement learning. At each simulated state $s'_{sim}$, the policy $\pi'$ outputs an activation $a'_{sim}$, while the fixed behavioral-cloning proxy $\pi_{BC}$ produces its reference activation $a'_{BC}$. Their discrepancy defines a supervised imitation loss $L_{BC} = \|a'_{sim} - a'_{BC}\|$, which helps preserve patient-specific activation characteristics. This term is added to the standard PPO loss (29) with coefficient $\alpha_{BC}$, which linearly decays from 0.2 to 0 over the first 5 M steps so that the contribution of BC supervision gradually diminishes as reinforcement learning progresses. In parallel, the GAIL discriminator $D_\theta$ compares a batch of generated pairs ($s'_{sim}$, $a'_{sim}$) with real patient pairs ($s'_{ref}$, $a'_{ref}$) sampled from the reference clip, learning to distinguish the two underlying distributions. Its output is converted into a per-timestep intrinsic imitation reward $r_{\text{in}}$, which is combined with the physics-based external reward during PPO updates:

$$r_{\text{in}}(s, a) = -\log\left[1 - D_\theta\left((s_{sim}, a_{sim}), (s_{ref}, a_{ref})\right)\right]$$

Meanwhile, the physics-based controller provides an external reward $r_{ex} = r'$. PPO then optimizes $\pi'$ using the combined reward $r_{\text{tot}} = r_{\text{ex}} + r_{\text{in}}$, producing an update step informed jointly by physical feasibility and imitation fidelity. The final parameter update further incorporates the behavioral-cloning supervision term weighted by $\alpha_{\text{BC}}$, yielding a controller that retains patient-specific activation characteristics while adapting robustly to the rehabilitation task.

The policy and discriminator networks are feedforward networks (policy 3×256; value and discriminator 2×128, ReLU). We linearly decay PPO/GAIL learning rates ($2\times10^{-5}/1\times10^{-5}$) and weight external/internal rewards at 0.9/0.1.

**Task-Conditioned Rehabilitation Evaluation**

This study presents two training scenarios involving typical tasks for patients in daily life and rehabilitation training: slope ascent and stair climbing. Curriculum Learning (CL) techniques (50) are employed to adjust task difficulty, such as varying slope angles and step heights, and accelerate convergence by adapting learning speed. We apply CL to set five difficulty levels for each task. Difficulty is promoted when the cumulative reward exceeds a threshold of 280 (90% of the patient's flat-ground baseline reward) and at least 500 episodes have been completed.

To monitor the training process, we track GAIL-related quantities, including the internal imitation reward, discriminator output, and discriminator loss. These quantities are used as model-internal training diagnostics rather than independent measures of prediction accuracy. In particular, the discriminator output $p \in [0, 1]$ reflects the adversarial distinguishability between generated and reference activation patterns. Following standard GAIL interpretation, values close to 0.5 indicate indistinguishability between generated and reference activation patterns. We therefore define an activation-signal imitation similarity score as $S = 1 - |p - 0.5| / 0.5$ where higher values indicate closer pattern-level alignment.

To quantitatively evaluate the quality of generated motion predictions, we employ two complementary metrics that assess joint-angle deviation and waveform-level trajectory correspondence, respectively. First, posture prediction accuracy quantifies the similarity between generated and real patient motions using joint-angle deviations aggregated over time and across articulated joints, computed based on the joint-angle imitation formulation used during training. The full mathematical definitions are provided in Supplementary Note S1. Second, joint profile PCC computes the Pearson correlation coefficient between generated and measured joint-angle trajectories after time normalization, providing a complementary waveform-level assessment.

*Task I - Slope Ascent*

We evaluate performance while maintaining the patient's nominal speed and step length across a five-level curriculum from flat ground to a final 1:12 incline. The ramp is 9 m long with elevations 0, 0.1875, 0.375, 0.5625, 0.75 m. Slope is computed as: slope = (height difference/horizontal distance) × 100%. In real life, outdoor ramps for disabled individuals typically have slopes ranging from 1:12 to 1:20.

*Task II - Stair Climbing*

We train and evaluate the patient-specific controller to climb stairs without preset speed targets, following a five-level schedule with step heights of 0, 3.75, 7.5, 11.25, 15 cm (typical real-world steps: 10–15 cm height, 30 cm depth). This task tests the method on clinically challenging rehabilitation scenarios. Prior work shows that example-guided DRL often struggles in stair environments because of large mismatches between reference actions and task objectives (51, 52). To mitigate this, we introduce a minimal terrain cue: if a step is detected within 5 cm ahead of the foot, a leg-raise action receives reward 1 (otherwise 0; if no step is ahead, reward 1). The cue only facilitates initial learning and is disabled once mastery is achieved, preserving final motion predictions (51). Task environments are visualized in Fig. S9. Integrated metrics of curriculum mastery and movement quality translate into patient-specific task and difficulty recommendations, supporting individualized and task-appropriate rehabilitation planning.

**Clinician Use of Simulated Rehabilitation Videos**

For each enrolled patient in the model-assisted group, the generative model produced short, patient-specific simulation videos of slope ascent and stair climbing at predefined difficulty levels, using only the 20 m level-ground walking trial collected at the start of that session as input. Rehabilitation physicians or physical therapists involved in rehabilitation planning necessarily had access to these simulated videos only for participants in the model-assisted group and were therefore not blinded to group allocation. Outcome assessments (FMA-LE) were performed by independent therapists who were blinded to group assignment. Patients were not explicitly informed of their allocation.

Rehabilitation physicians or physical therapists review these videos alongside routine clinical information to judge whether a given task level was likely to be feasible, to anticipate potential compensatory movements, and to adjust the starting difficulty or progression speed of conventional lower-limb training. As part of rehabilitation planning, these predictions informed clinician judgment regarding (i) whether slope ascent or stair climbing should be introduced for a given patient at a given stage, (ii) the initial difficulty level at which these tasks should be practiced, and (iii) whether progression to higher difficulty should be advanced or delayed over successive sessions. The system does not generate explicit prescriptions or "safe/unsafe" labels; it serves purely as a visual decision-support aid. All final decisions regarding task selection, intensity, and progression follow local clinical guidelines and the independent judgment of the treating clinician.

## Acknowledgments

We thank Y. Lu, R. Lv, B. Chen, J. Xue, and X. Li for their assistance with participant recruitment and data collection, and for fruitful discussions of data.

**Funding:**

Y.P. acknowledges funding from the National Key Clinical Specialty Capacity-Building Project (grant No. XKB2023A2001). S.G. acknowledges funding from the National Natural Science

Foundation of China (grant No. 62571016), and Beihang University (grants No.KG161250 and ZG16S2103), which supported and hosted the core research activities reported in this work.

**Author contributions:**

Conceptualization: Y.D. and S.G. Methodology: Y.D. and R.Z. Validation: Y.D. and R.Z. Visualization: Y.Z., Y.W., and W.Y. Investigation: R.Z., W.Y., J.Ch., and X.C. Resources: R.X., Y.C., Q.L., J.Ca., T.L., and Y.P. Supervision: C.T., Y.P., and S.G. Writing—original draft: Y.D., C.T., and S.G. Writing—review & editing: H.Z., X.G., A.N., P.S., and S.G. Project administration: C.T., Y.P., and S.G.

**Competing interests:**

The authors declare they have no competing interests.

**Data, code, and materials availability:**

Data and code associated with this study are archived on Zenodo at https://doi.org/10.5281/zenodo.21463294. The archive includes de-identified gait sequences from 19 healthy individuals and 21 patients with stroke, processed kinematic features used for model training and evaluation, and all code necessary to reproduce the main experiments reported in this study. As an additional development resource, the corresponding GitHub repository is available at https://github.com/YanningDai/StrokeAvatar. No new materials were created for this study.

**Computational environment and resources**

All simulation scenes and 3D assets are built and rendered in Unity 3D; rigid-body dynamics are computed by NVIDIA PhysX. Reinforcement learning and control interfaced with the simulator via Unity ML-Agents (Release 20), with training implemented in PyTorch 1.8.0 (CUDA 11.1) on Python 3.8.13. Experiments run on a workstation equipped with an NVIDIA GeForce RTX 3090 (24 GB VRAM) and an Intel i7-12700 (12 cores, 2.10 GHz); the control/physics loop operated at 30 Hz. Detailed hyperparameters are provided in Table S12.

# Supplementary Materials for

## Wearable-informed generative digital avatars predict task-conditioned post-stroke locomotion

Yanning Dai *et al.*

Corresponding author: Chenyu Tang, ctang@pku.edu.cn; Yu Pan, panyu@btch.edu.cn;
Shuo Gao, shuo_gao@buaa.edu.cn

**This PDF file includes:**

Supplementary Notes S1 to S2
Figures S1 to S9
Tables S1 to S12
Legends for movies S1

**Other Supplementary Materials for this manuscript include the following:**

Movie S1

## Supplementary Note S1. Reinforcement learning reward formulation and implementation

To ensure reproducibility and clarity, this note details the exact formulation of the reinforcement learning reward function $r_t$, including all component terms, coefficients, and implementation settings used in training. The formulation of $r_t$ is as follows:

$$r_t = w_1 r_{goal} + w_2 r_{imitation} - p_{physics}$$

where $w_1$ and $w_2$ are both set to 0.5. $r_{goal}$ is further decomposed into 4 terms that reward the character for matching global parameters of a selected clip from the dataset. These terms include: walking speed reward $r_{g(velocity)}$, step length reward $r_{g(length)}$, direction reward $r_{g(forward)}$, and root node position reward $r_{g(hip)}$ (comprising root node height and lateral sway):

$$r_{goal} = g_1\, r_{g(speed)} + g_2\, r_{g(length)} + g_3\, r_{g(forward)} + g_4\, r_{g(hip)}$$

Weights for each component are determined based on preliminary calibration experiments, in which the typical magnitude ranges of different error terms are estimated. The weights are then chosen to normalize these components to comparable scales, preventing any single term from dominating the overall reward; in this study, they are set to $\{0.5, 0.1, 0.2, 0.2\}$. The calculations for each component are as follows:

$$r_{g(speed)} = exp\left[-2.5\left(v_{goal} - \sum_{i}^{n} v_i^{hip}\right)^2\right], \quad n = 100$$

$$r_{g(length)} = exp\left[-\left(1 - \max(0, L_t)/L_{goal}\right)^2\right]$$

$$r_{g(forward)} = exp\left[-\left(1 - \left(p_t^{hip} - p_0^{hip}\right)\cdot\left(p_{t(goal)}^{hip} - p_0^{hip}\right)\right)^2\right]$$

$$r_{g(hip)} = 0.5 \times exp\left[-40\left(h_t^{hip} - h_{goal}^{hip}\right)^2\right] + 0.5 \times exp\left[-40\left(l_t^{hip}/l_{goal}^{hip}\right)^2\right]$$

where $v_{goal}$ is the target speed, and $v_{hip}$ is the current speed of the hip, both determined by the average speed over the past $n$ time steps, which is set to 100 in this study. $L_t$ and $L_{goal}$ are the current and target step lengths, respectively, calculated as the root node's forward distance during the latest walking cycle. The cycle is defined by the interval between two instances of the same phase value $\phi$. $p_t^{hip}$ is the current root node position, and $p_{t\,(goal)}^{hip}$ is the target root node position; the difference between them and the initial position $p_0^{hip}$ indicates the forward direction. $h_t^{hip}$ and $h_{goal}^{hip}$ are the current and target root node heights, respectively, while $l_t^{hip}$ and $l_{goal}^{hip}$ represent the current and target lateral sway of the root node.

$r_{imitation}$ assesses the consistency between the current model and target clip in joint angles, linear velocities, angular velocities, as well as the CoM position and endpoint positions (i.e., spine top and foot):

$$r_{imitation} = i_1\, r_{i(angle)} + i_2\, r_{i(Avelocity)} + i_3\, r_{i(velocity)} + i_4\, r_{i(end)} + i_5\, r_{i(com)}$$

where weights for each reward are $\{0.55, 0.15, 0.05, 0.15, 0.10\}$. The angle imitation reward $r_{i(angle)}$ measures the average imitation similarity of the rotation angles of each body part:

$$r_{i(angle)} = \frac{1}{6}\sum_{j} exp\left[-\frac{1}{\sigma_j^2}\|\hat{q}^{\,j}_{t+1} \circ q^{j}_{t+1}\|^2\right]$$

where $q_t^j$ represents the current local rotation quaternion, $\hat{q}_t^j$ represents the target local rotation quaternion, and $\circ$ denotes the quaternion difference operation. $\|q\|$ represents the scalar rotation radians around the axis of the quaternion. $\sigma_j$ is set to 2 for the left and right knee joints and 1 for other joints. Note that the root node angle is referenced in the global coordinate system, while the angles of other body segments are referenced in their respective parent node's coordinate system.

The remaining four terms in $r_{imitation}$ are three-dimensional vectors, calculated by:

$$r = \sum_e \exp\left[-\frac{1}{\sigma_e^2}\ \|\ \hat{p}^e_{t+1} - p^e_{t+1}\ \|^2\right]$$

where $p_t^e$ represents the current angular velocity, linear velocity, and relative position in the simulation, $\hat{p}_t^e$ represents the corresponding target information, and $\|p\|$ denotes the magnitude of the vector. $\sigma_e$ is set to 0.15 for the CoM, 0.25 for the left and right foot endpoints, and 0.18 for the spine endpoint, 1 for spine and thigh joint velocities, and 2 for knee joint velocities. $p_{physics}$ refers to penalties for foot-dragging and floating:

$$p_{physics} = p_{float} + p_{drag}$$

where $p_{drag}$ penalizes sliding when the feet contact the ground:

$$p_{drag} = \sum_{k \in feet} f^k_{t\text{-}1}\, f^k_t\ \|\ p^k_t - p^k_{t\text{-}1}\|^2$$

This penalty is activated only when a foot is in ground contact and exhibits sustained lateral or forward sliding in the simulator, reflecting violations of contact and friction constraints. It does not penalize reduced toe clearance, prolonged stance, or other swing-phase compensatory patterns commonly observed in post-stroke gait, which are preserved through imitation-based motor-control rewards. Here, feet = [left foot, right foot], and $f$ denotes the contact status of each foot with the ground: 1 if the foot is in contact, and 0 otherwise. In addition, if any body part other than the feet touches the ground, the step reward is set to 0, and the training round terminates. $p_{float}$ applies a slight negative penalty (– 0.005) when both feet are off the ground:

$$p_{float} = -0.005 \prod_{k \in feet} \left(1 - f^k_t\right)$$

**Supplementary Note S2. Related works in human motion generation techniques**

Human motion generation techniques can be broadly categorized into data-driven, physics-driven, and data–physics hybrid paradigms. Each paradigm reflects a different philosophy of modeling human movement—either learning motion patterns from data, constraining them through physical priors, or integrating both to achieve balance between realism, feasibility, and adaptability.

***- Data-Driven Models***

Data-driven approaches learn statistical regularities of human motion from large datasets, relying on examples rather than explicit physical modeling. Early methods constructed motion graphs or motion grammars by segmenting and recombining existing motion clips to synthesize new sequences with smooth transitions (53, 54). With the advancement of machine learning, these handcrafted structures were gradually replaced by sequence learning and generative paradigms, where temporal dependencies are modeled through recurrent or attention-based architectures (55–57). Recent trends also include generative models (58, 59) and large language models (LLMs) (60) that capture higher-level motion style or intention, enabling automatic synthesis of diverse actions such as walking, running, or jumping.

These data-driven models excel in reproducing realistic movement dynamics and capturing individual variability (61), which is valuable for clinical datasets containing heterogeneous motion styles. However, their lack of explicit physical constraints often leads to unrealistic joint configurations, foot sliding, or loss of balance when applied to unseen terrains or pathological motions. In rehabilitation contexts, such limitations reduce their reliability for predicting physically feasible recovery trajectories.

***- Physics-Driven Models***

Physics-driven approaches, in contrast, rely on biomechanical or dynamic modeling to generate motion through physically interpretable parameters such as joint torques, muscle activation, or reaction forces (62, 63). Biomechanical formulations represent muscles, tendons, and joints with parameterized elements (e.g., Hill-type muscles) and use forward or inverse dynamics to map muscle activations to joint torques and whole-body posture. They can estimate internal loads and reveal mechanisms of impairment such as weakness, spasticity, and abnormal co-activation, which is valuable for hypothesis-driven rehabilitation planning (64, 65). However, subject-specific calibration is demanding, parameters are uncertain in neurological populations, and whole-body task-adaptive simulations often require simplifications that limit scalability (63, 64). Open-source toolchains (e.g., AnyBody, OpenSim) enable personalization, but per-patient calibration remains labor- and time-intensive (66).

A complementary line of work treats the body as a linked rigid-body system under Newtonian dynamics and contact/friction constraints. Classical controllers solve trajectory optimization or inverse dynamics and then stabilize motion with feedback or hierarchical schemes to realize user-specified goals such as walking speed, step length, or heading direction (67, 68). More recently, deep reinforcement learning (DRL) has been used to learn control policies directly in physics simulators with rewards that encode stability, energy economy, torque smoothness, and contact fidelity (51, 69). Within DRL, methods without reference explore policies from scratch and can exhibit diverse yet occasionally unnatural or inefficient strategies under prolonged simulations (70). Methods with reference incorporate motion clips or capture data into the objective (e.g., imitation losses or reference-matching rewards) to encourage naturalistic coordination while retaining physical feasibility and environmental responsiveness (31, 71). Despite these advances, the underdetermined nature of human movement means many physically valid solutions exist; without additional priors, controllers may drift toward atypical gaits and still demand substantial tuning for patient-specific deployment (69). These limitations motivate data–physics hybrids that introduce learned priors while preserving hard physical constraints.

***- Data-Physics Hybrid-Driven Models***

Data–physics hybrid approaches seek a balance between preserving data-driven self-motion patterns and enforcing physical feasibility, and have therefore been explored for human motion generation and control (72–74). Rather than committing to a single paradigm, hybrid controllers combine learned priors with dynamics constraints through several recurring strategies. One line alternates between modes: a short burst of physics simulation provides a state or terminal posture that is then matched to examples in a motion database, with transitions synthesized by data-driven modules to mimic responses to external impacts and pushes (72, 75). Another line performs joint learning of data-driven and physics-driven components for the *same* action, using predictions from the former to shape torque targets or trajectories for the latter, thereby improving robustness to disturbances while retaining imitation fidelity (76). A sequential design first uses a data-driven generator to propose reference motions from user inputs and then refines these proposals with a physics controller—often through reinforcement learning—to guarantee contact handling and dynamic stability, though the action space is implicitly restricted by what exists in the library (73, 74). A different integration fuses outputs at the decision level, where motion and physics controllers run in parallel, and their recommendations are combined to enhance environmental interaction, at the risk of accumulating errors from the data-driven branch (77). Frame-wise fusion has also been proposed: a data-driven predictor (e.g., a recurrent policy) emits the next action, which is concatenated with physics-informed input to form the final control signal; this limits drift but requires substantial data and careful tuning, making individualized analysis challenging (78, 79).

Imitation-based reinforcement learning provides a particularly effective instantiation of the hybrid idea. Generative Adversarial Imitation Learning (GAIL) distinguishes reference versus generated motions while optimizing for task completion, bringing data realism and physical objectives into a single training loop (30). Although GAIL can operate with minimal per-patient data, its feature-based discriminator may over- or under-penalize stylistic deviations, creating discrepancies between generated and original action patterns and necessitating additional constraints or rewards in practice (52).

This study integrates behavioral cloning (BC) with GAIL within a physics-based reinforcement learning framework, aiming to couple interactive adaptability with predictive accuracy in heterogeneous, previously unseen terrains while leveraging limited individual data. Motion posture sequences serve as reference signals for the reinforcement learning component, whereas derived motor activation patterns provide a physiologically grounded, control-level imitation signal for the data-driven module. By constraining imitation at the level of control activation rather than posture alone, this formulation facilitates task generalization while preserving task feasibility and clinically meaningful kinematics.

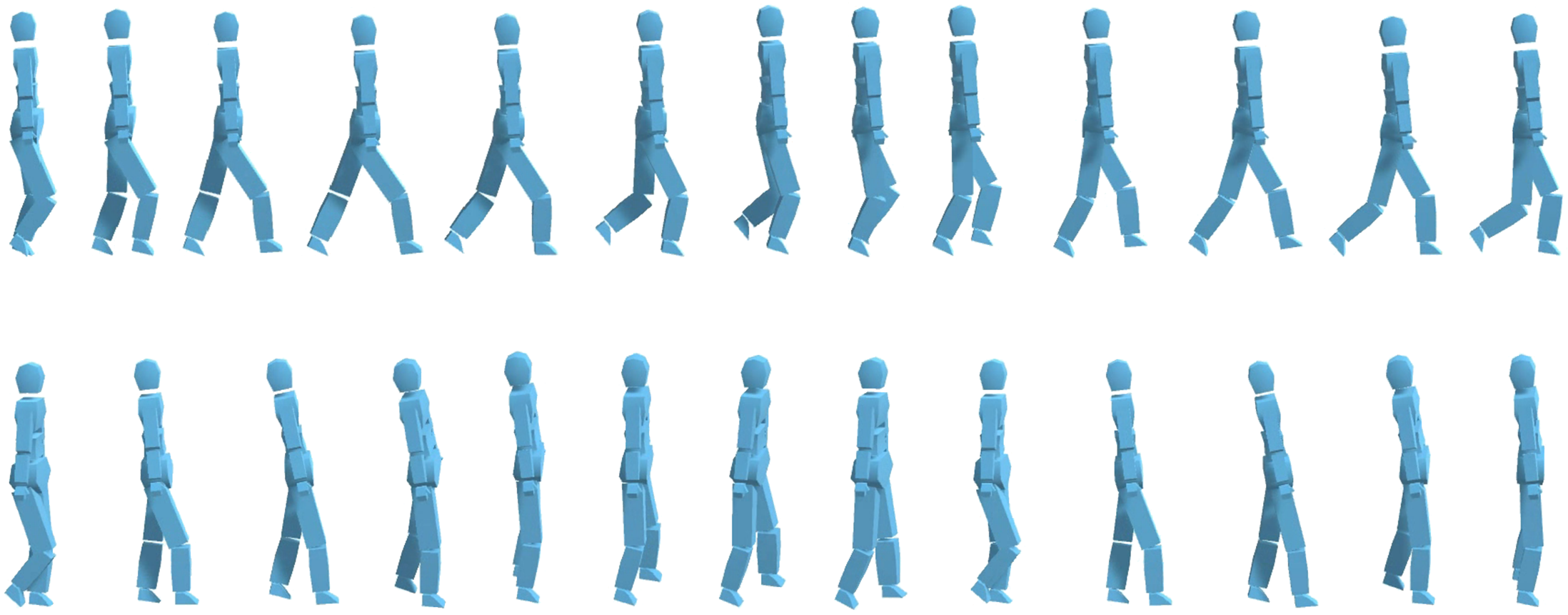

**Fig. S1. Reconstructed gait sequences with the physical avatar.** Reconstructed gait sequences are shown for two representative subjects—a healthy individual and a stroke patient—generated using the physical avatar model. The healthy gait exhibits a step length of 1.61 m and a forward speed of 1.24 m/s, while the patient's gait shows 0.80 m and 0.71 m/s, respectively. Here, step length is defined over one full gait cycle (i.e., consecutive contacts of the same foot). An obvious asymmetry can be observed in the patient's gait cycle. On the impaired side, the knee fails to flex properly, and forward progression relies on hip elevation of the contralateral limb, resulting in a circumduction pattern. In rhythm, the stance phase of the healthy limb is markedly longer than that of the affected side.

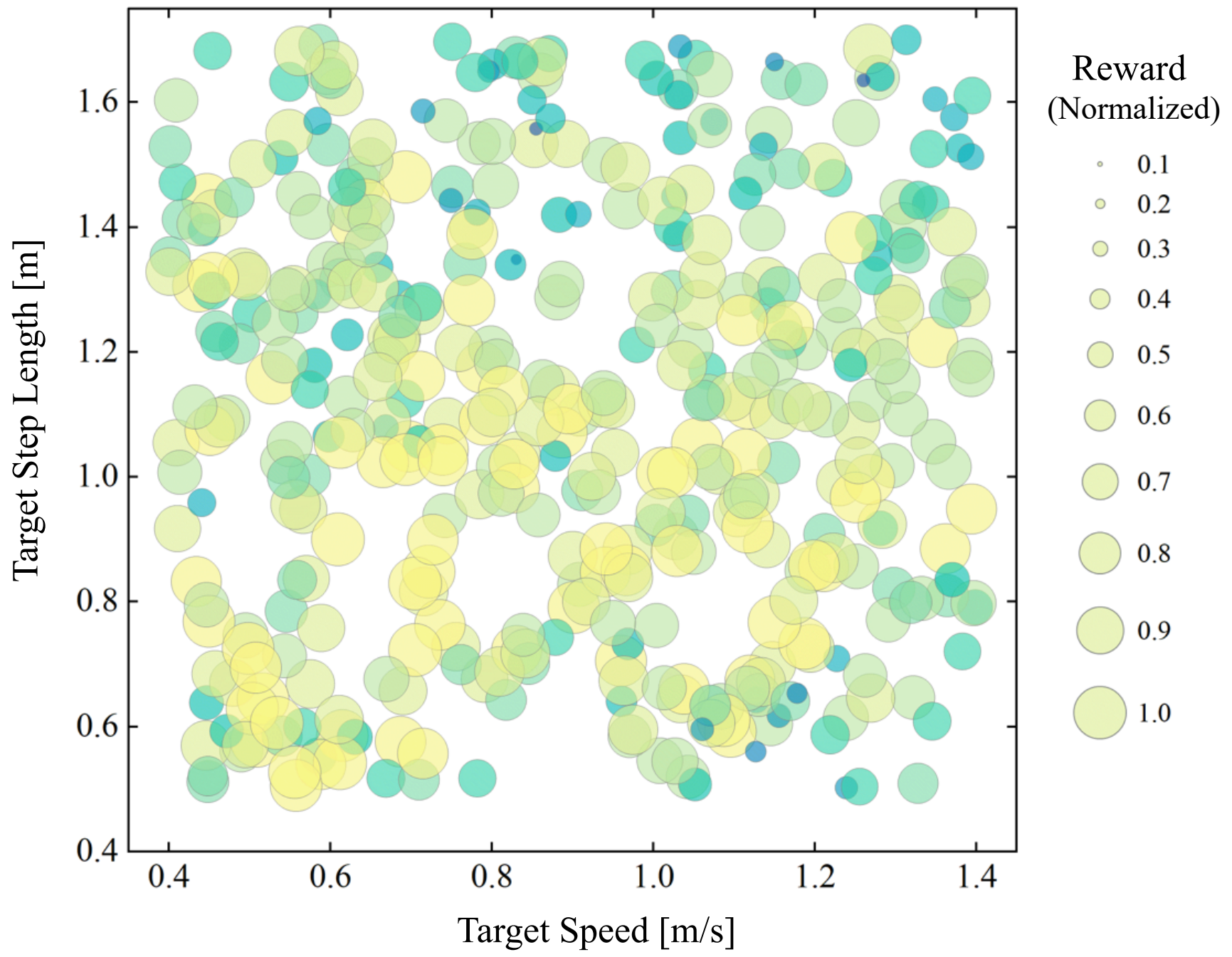

**Fig. S2. Sensitivity of the Healthy Motion Atlas to target speed and step length.** Scatter of 400 target pairs (speed, step length) uniformly sampled within the trained domain; each pair was evaluated with 150 rollouts, and the mean normalized reward (0–1) is shown by color/size. Rewards are largely uniform across the grid, indicating stable limit cycles over most targets. A mild decline appears at the fastest speeds and at sparsely sampled extremes of step length, consistent with increased dynamic difficulty and thinner prior coverage in those regions.

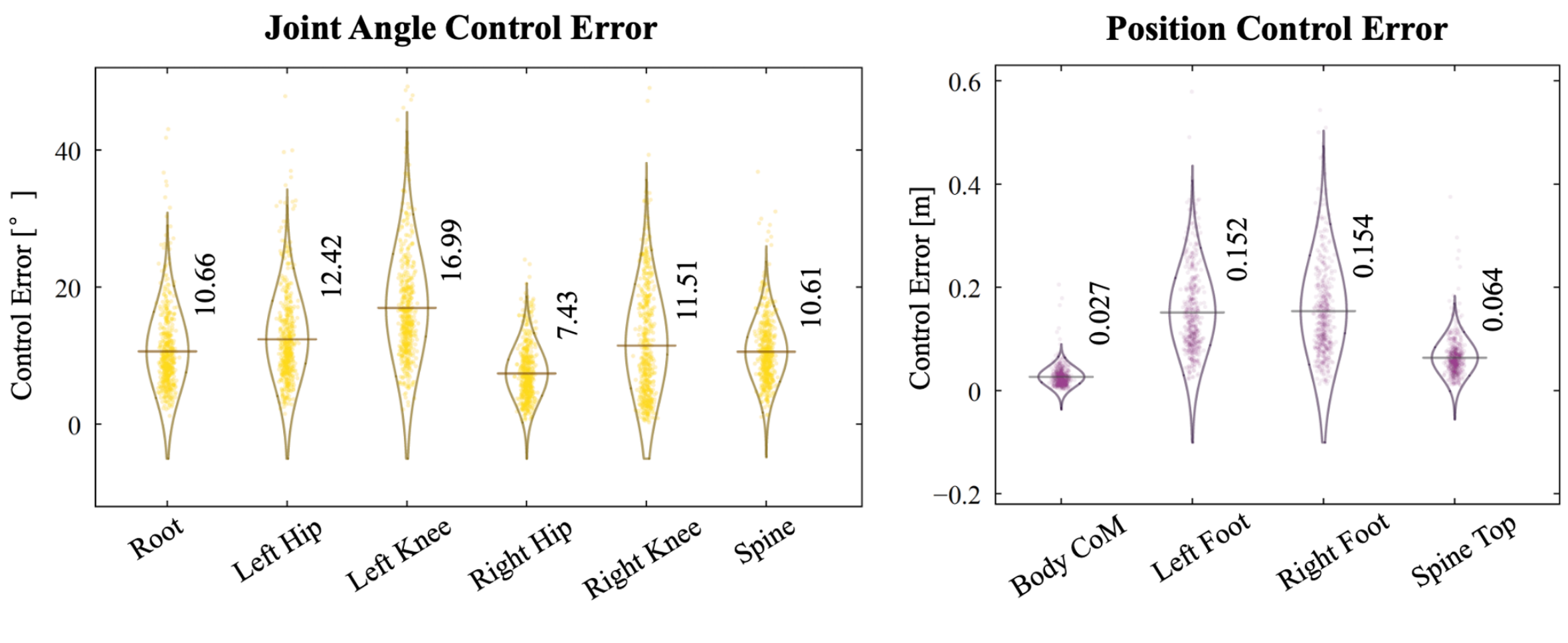

(A) (B)

**Fig. S3. Joint- and endpoint-level control errors for Healthy Motion Atlas.** Across 500 randomized trials with targets uniformly sampled over speed and step length, the Healthy Motion Atlas yields a mean single-axis joint-angle error of 11.61°, a CoM position error of 2.69 cm, and an end-effector position error of 12.32 cm. Violin plots show the error distributions: (**A**) joint-angle RMS errors for root, left/right hip, left/right knee, and spine segments (means annotated inside violins); knees exhibit slightly larger deviations, reflecting their role as multi-link regulators during stance-to-swing transitions and distal error accumulation. (**B**) CoM and endpoint positional root mean square (RMS) errors remain low, indicating that balance is well maintained while distal segments adjust.

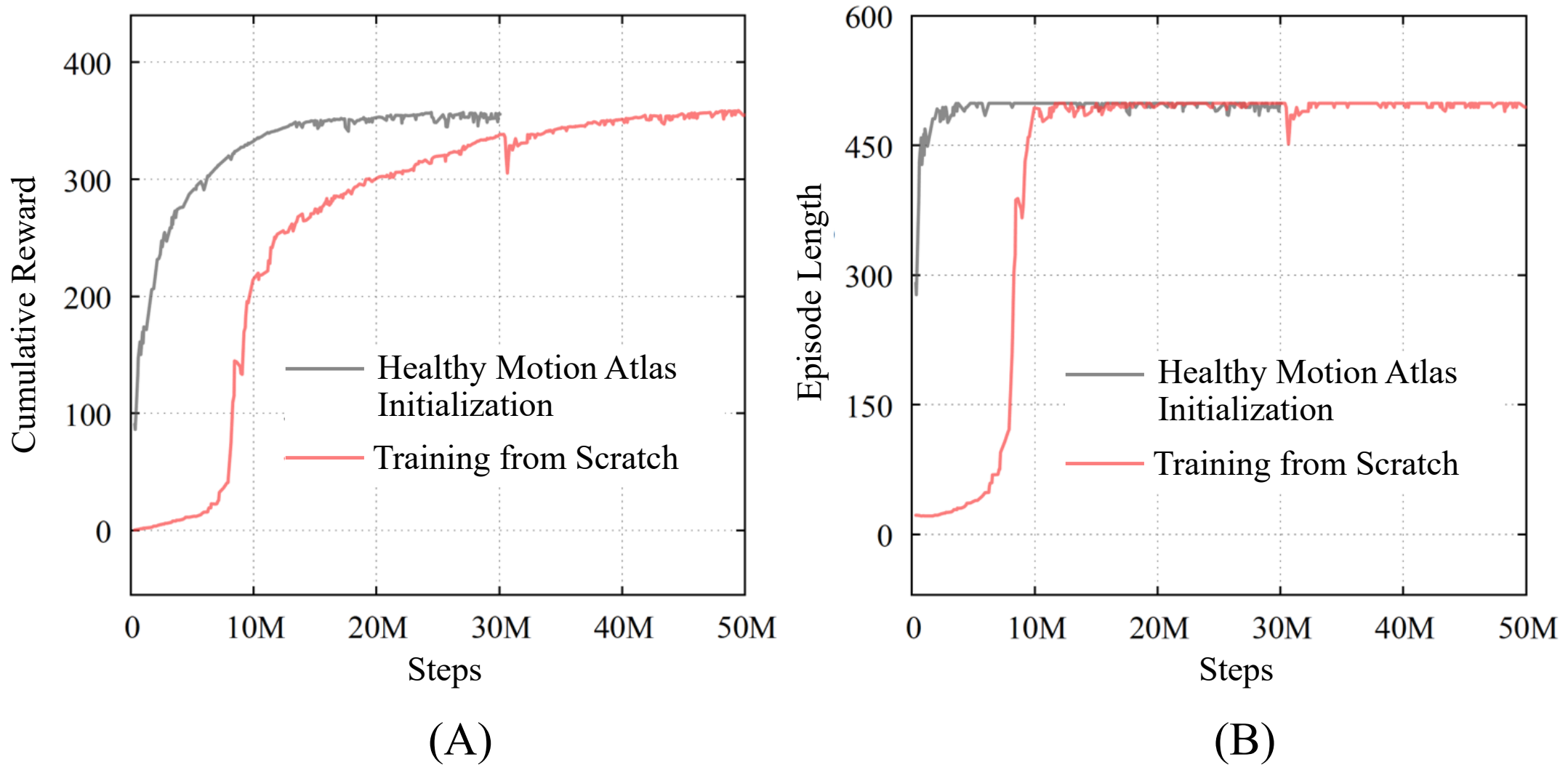

**Fig. S4. Effect of atlas initialization on training efficiency.** Comparison of reinforcement-learning training with Healthy Motion Atlas initialization (gray) versus training from scratch (red). (**A**) Cumulative-reward learning curves. (**B**) Sequence-length learning curves. Atlas initialization reaches stable 500-step episodes after ~7 M timesteps (≈3 h), whereas scratch training requires ~15.7 M timesteps (≈8 h), reducing wall-clock time to ~29.8% and accelerating convergence.

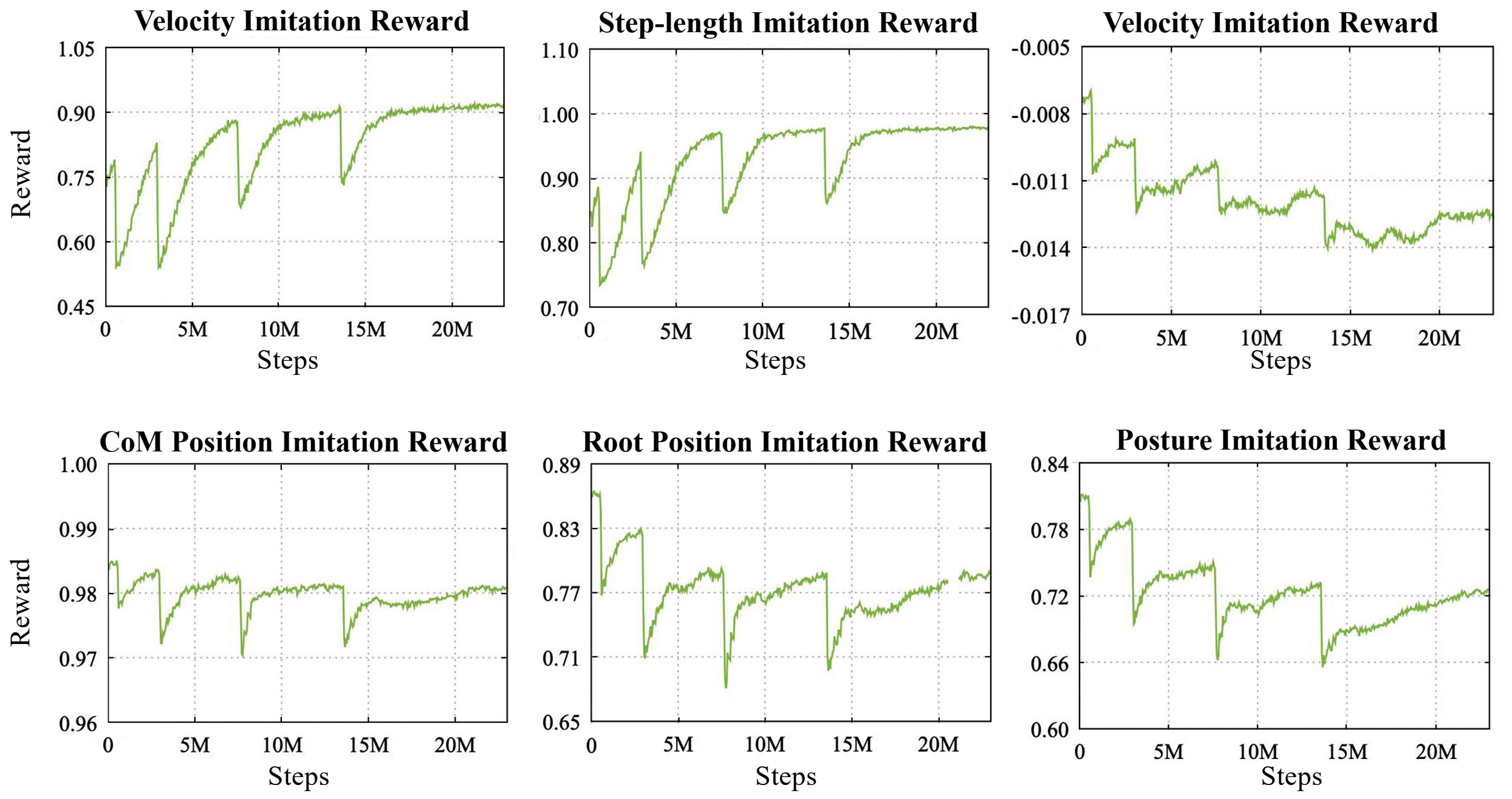

(A)

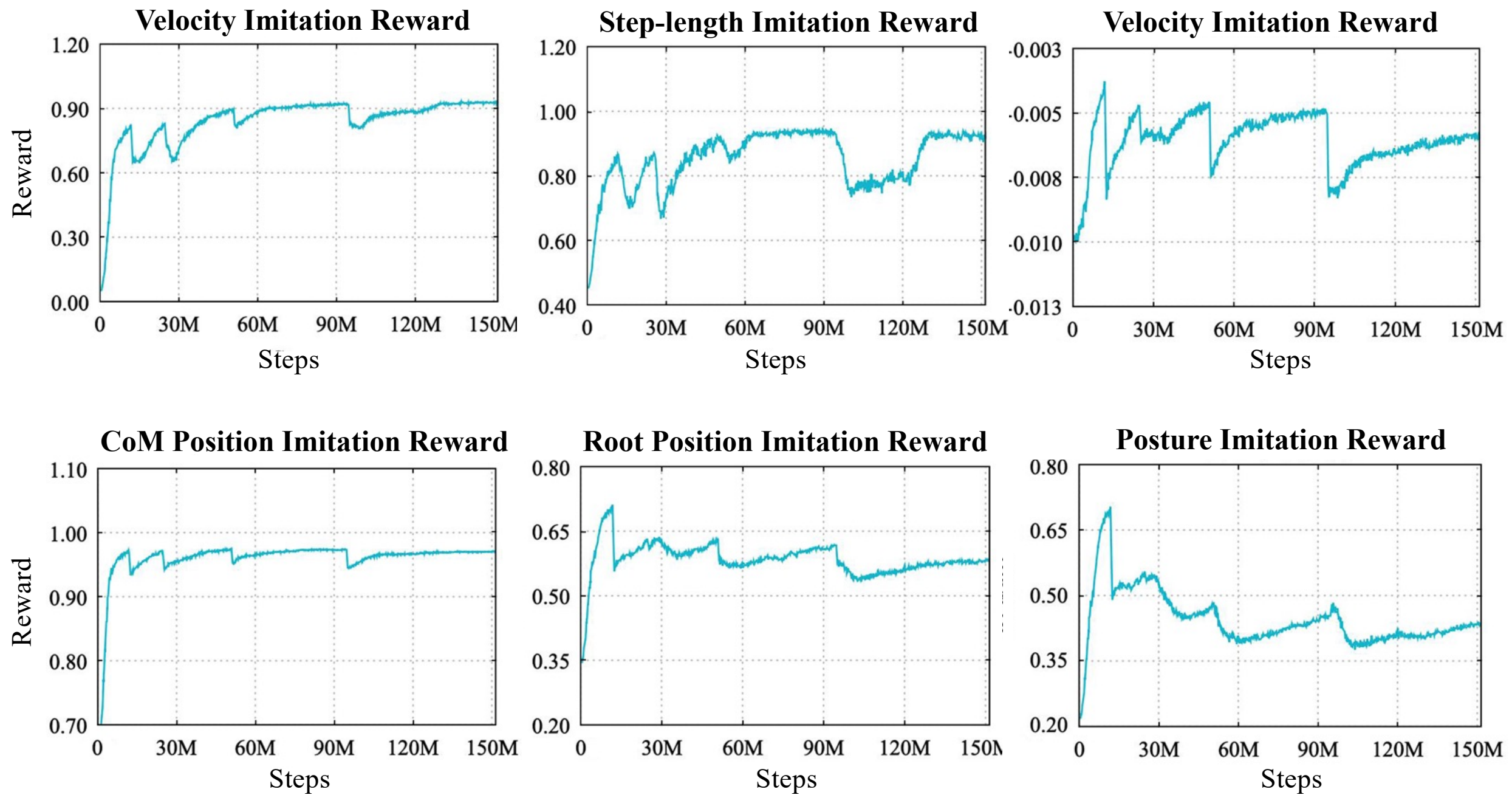

(B)

**Fig. S5. Training curves of reward components for terrain-adaptive locomotion prediction.** Six components are tracked during training: speed similarity, step-length similarity, physical feasibility, center-of-mass (COM) position reward, root-joint position reward, and pose-level imitation reward. (**A**) *Slope Ascent:* training curves over 23 M environment steps. (**B**) *Stair Climbing:* training curves over 23 M environment steps.

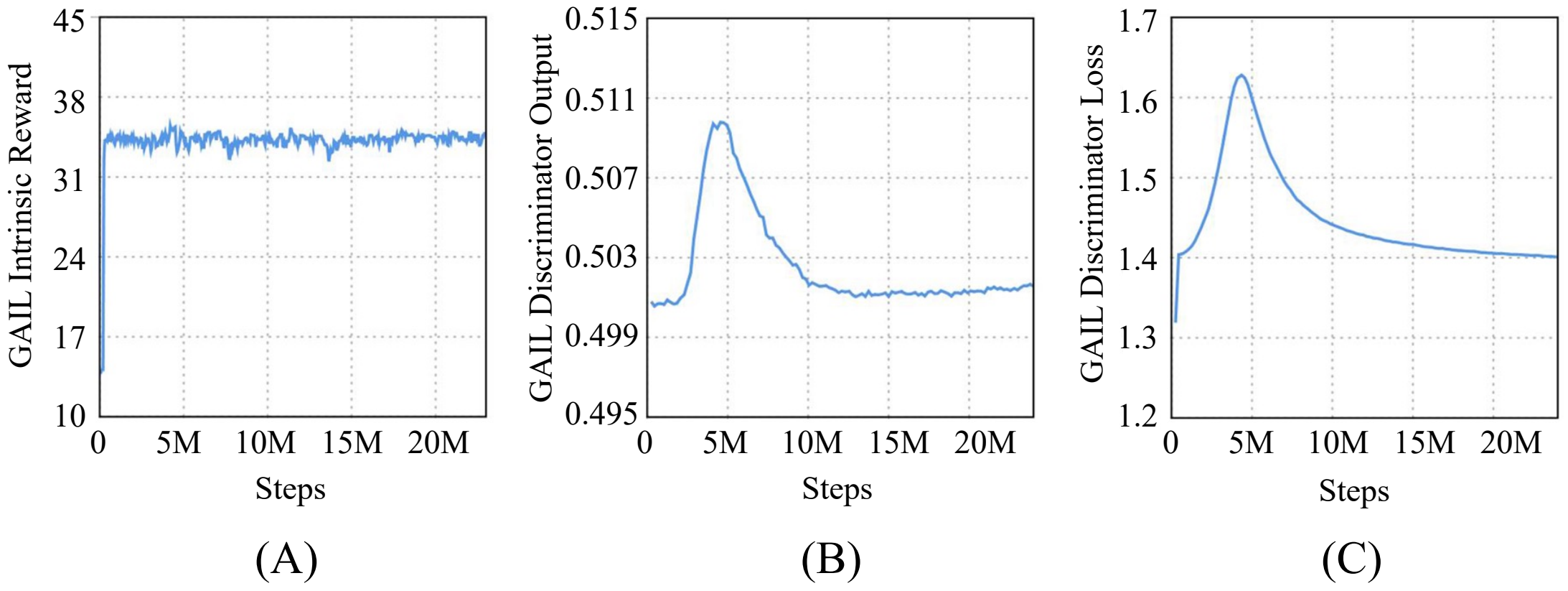

**Fig. S6. GAIL training dynamics for the slope-ascent curriculum.** (**A**) Internal reward versus environment steps, showing plateaus across curriculum stages. (**B**) Discriminator output versus environment steps. Convergence toward a stable band around 0.5 indicates that the generated activation patterns become indistinguishable from the patient's flat-ground reference. (**C**) Discriminator loss versus environment steps. The loss shows early-training rise and subsequent decay, stabilizing as the generator and discriminator reach equilibrium.

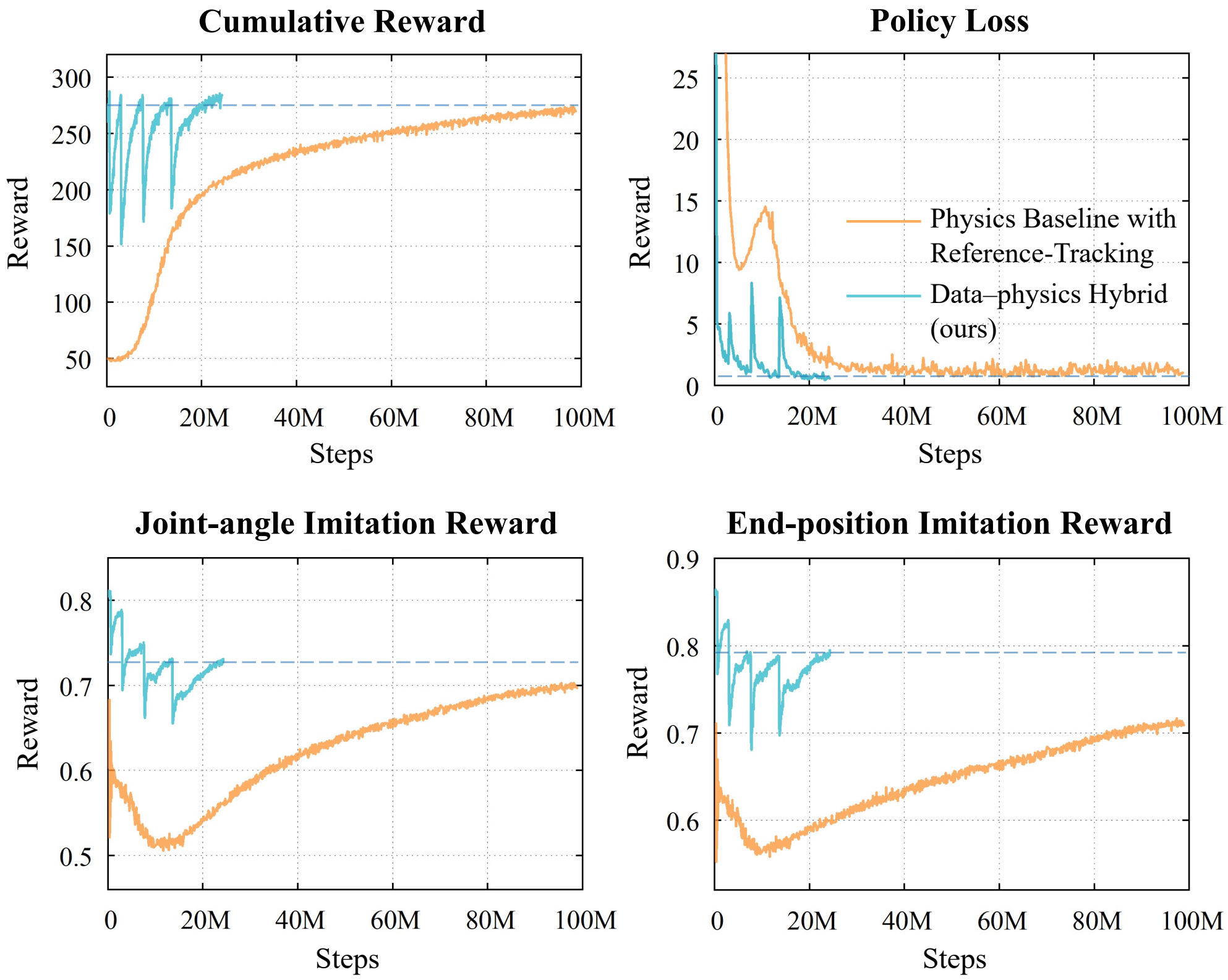

**Fig. S7. Baseline comparison.** Comparison between the proposed data–physics fusion controller and a physics baseline with reference-tracking on the 1:12 slope-climbing task. The plots display cumulative reward, policy network loss, joint-angle similarity reward, and distal end-effector position similarity reward over training. Our method converges by ≈23 M steps vs ≈90 M for the baseline, a 74.4% reduction in training time. Beyond efficiency, our method attains lower imitation loss, smoother policy transitions with a sustained rise in strategy reward, and higher motion fidelity, yielding +4.73% joint-angle accuracy and +12.10% endpoint position accuracy relative to the baseline.

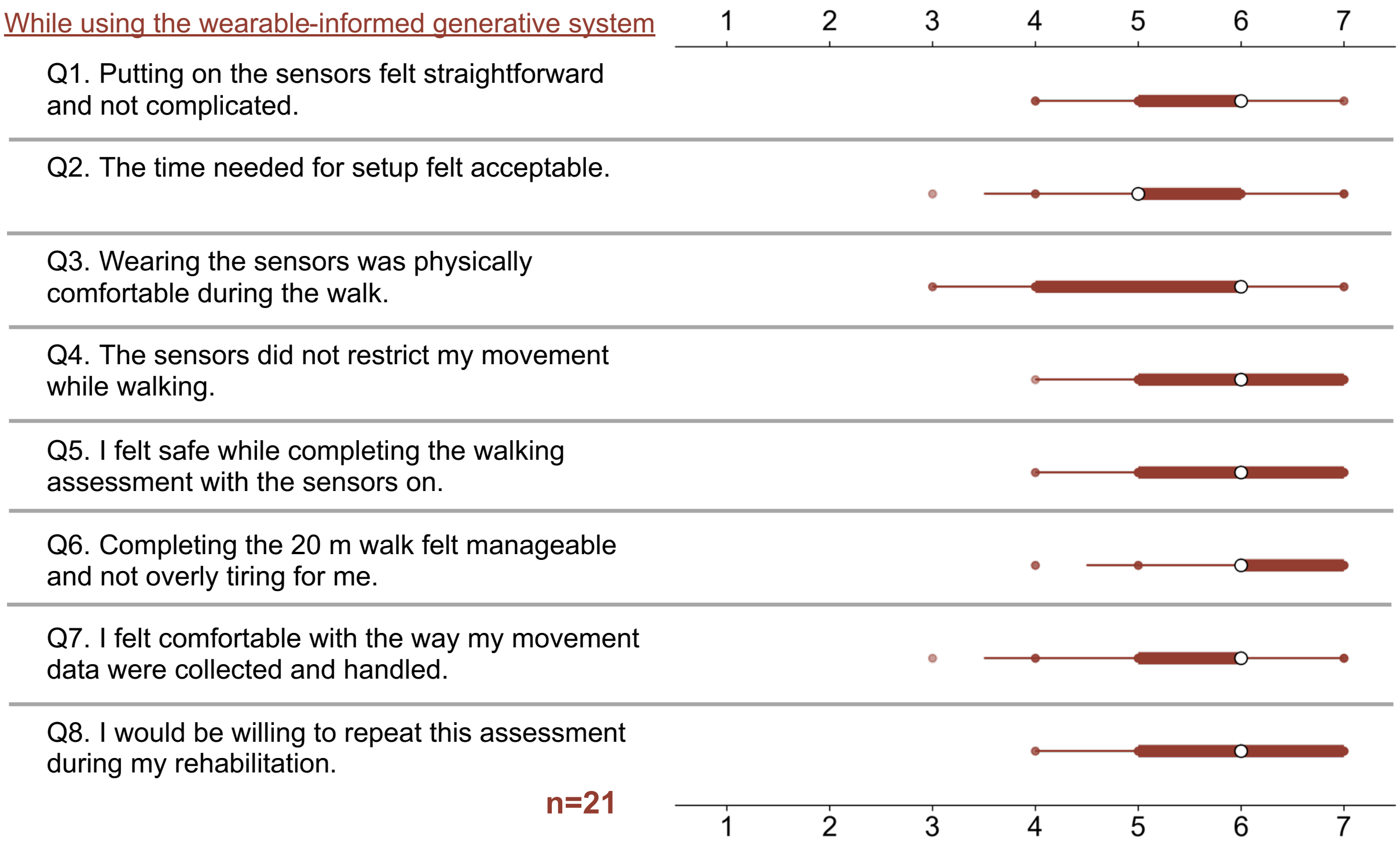

**Fig. S8. Patient-reported acceptability of the wearable-informed generative system.** Patient feedback is collected immediately after completing the wearable-based 20 m level-ground walking assessment. Responses were recorded using a 7-point Likert scale (1 = strongly disagree; 7 = strongly agree) across eight items assessing setup simplicity, time burden, physical comfort, movement restriction, perceived safety, exertion, data-handling comfort, and willingness to repeat the assessment during rehabilitation. For each item, individual responses are shown as semi-transparent dots, the interquartile range (IQR) is indicated by the thick horizontal bar, whiskers denote the full data range within 1.5×IQR, and the median is marked by a white circle. Higher scores indicate greater acceptability and lower perceived burden. n = 21 patients.

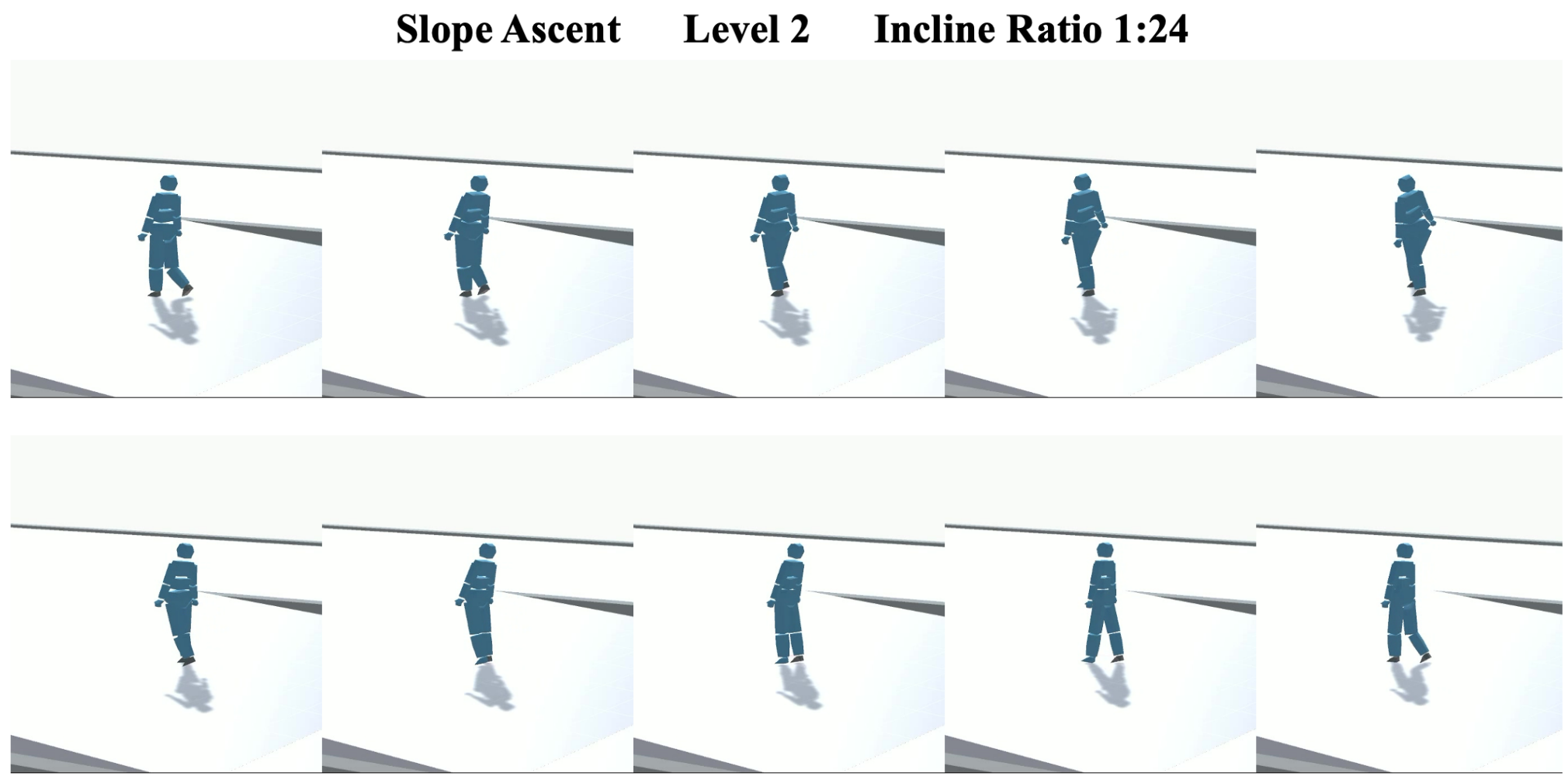

(A)

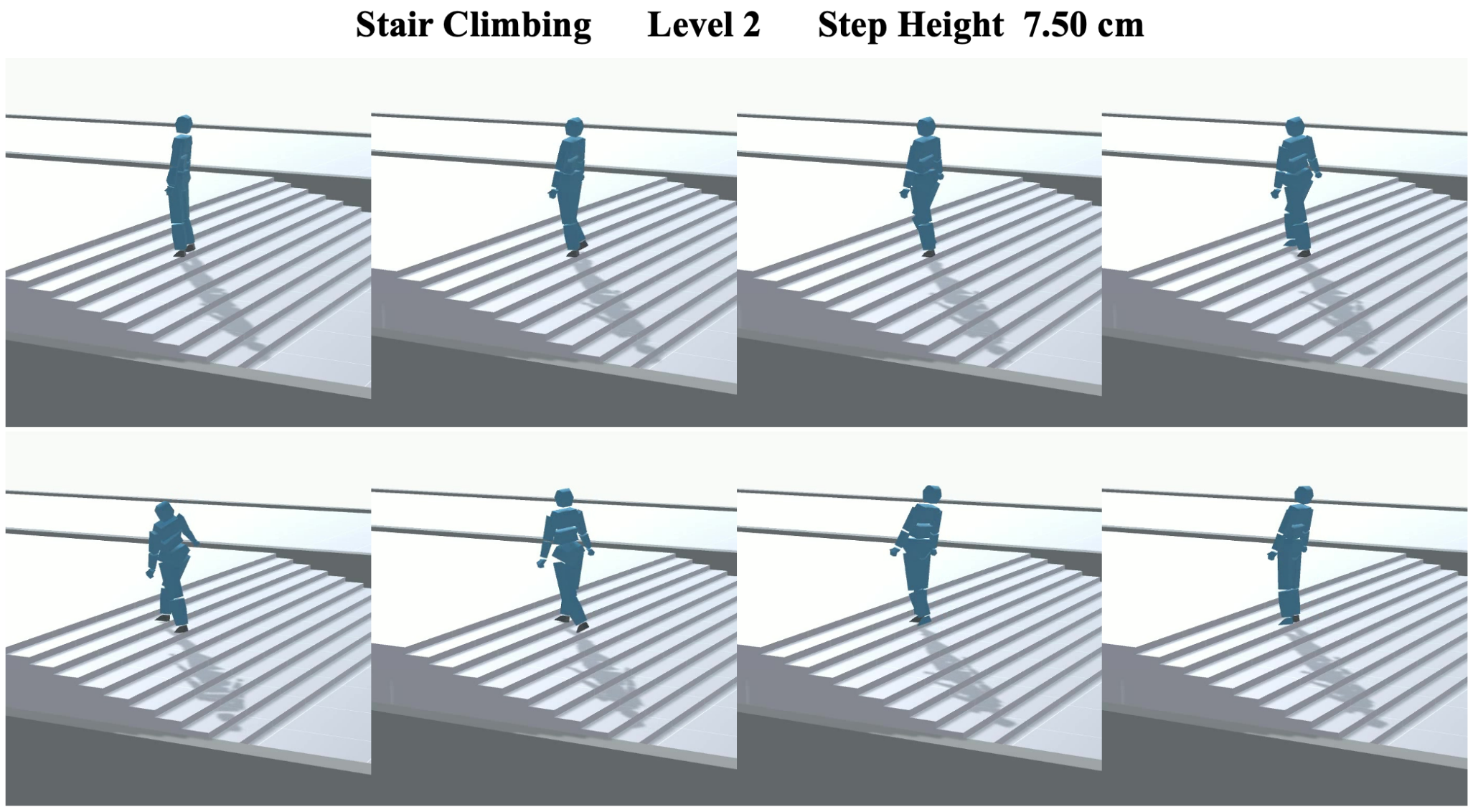

(B)

**Fig. S9. Task environments for terrain-adaptive rehabilitation.** Two locomotor rehabilitation scenarios are implemented as RL environments in Unity3D, with rigid-body dynamics computed by the PhysX physical engine (shared joint limits, contact, and friction). Both tasks expose the same observation–action interface as the patient-specific controller and use a five-level curriculum that progressively increases terrain demands. (**A**) *Slope Ascent*: a straight 9 m ramp whose elevation rises across levels to 1:12 (heights 0, 0.1875, 0.375, 0.5625, 0.75 m); target speed and step length are fixed—only the slope changes. (**B**) *Stair Climbing*: a staircase with 30 cm tread; step height progresses 0, 3.75, 7.50, 11.25, 15.00 cm. Panels show representative mid-curriculum frames.

**Table S1. Comparative overview of representative AI-assisted wearable approaches for motor assessment and rehabilitation.**

| Research | Data Source | AI Method | Disease | Main Objective | Clinical Utility |
|---|---|---|---|---|---|
| Islam, M. S. *et al.* (2023) (16) | Single camera | Regression (LightGBM) | PD | Remote quantification of PD severity | Objective proxy aligned with scale scoring |
| Yu, B. *et al.* (2024) (17) | 2 cameras | Anomaly Detection (confidence-based) | Stroke | Quantification of motion impairment | Objective proxy aligned with scale scoring |
| Lange, F. *et al.* (2025) (19) | Single camera | Unsupervised dimensionality reduction (Sparse PCA) | PD | Kinematic domain mapping of clinical scale dimensions | Dimensional refinement of clinical scales |
| Salomon, A. *et al.* (2024) (20) | Single lower-back IMU | Classification (Transformer-based) | PD | Detection and subtype labeling of gait abnormalities across scenarios | Automated, objective, and scenario-consistent assessment |
| Zhang, W. *et al.* (2024) (21) | 10 IMUs | Correlation analysis | PD | Identify quantitative gait biomarkers and disease-severity indicators for PD | Dimensional refinement of clinical scales |
| Lee, J. T. *et al.* (2025) (22) | Text-based case prompts | LLM with prompt engineering | Stroke | Assessment of accuracy, hallucination control, empathy, specificity, and actionability of LLMs for stroke care | Assist in diagnosis and rehabilitation guidance, but not for direct clinical decision-making |
| **This work** | 5 IMUs | Reinforcement learning (PPO+GAIL) | Stroke | Predict and visualize locomotion performance across heterogeneous rehabilitation tasks and difficulty levels | Provide dynamic, rich predictions across diverse scenarios to assist rehabilitation planning |

**Table S2. Hyperparameter configuration of the simulated human model.** Each segment's degrees of freedom (DoF) and motion limits (flexion/extension, abduction/adduction, rotation) are specified together with the capsule rigid-body size (length/radius) and segment mass. The pelvis serves as the root, the spine provides three rotational DoFs, and the left/right hips and knees are articulated joints for dynamic simulation.

| Joint | DoF | Flexion/ Extension | Abduction/ Adduction | Rotation | Rigid Body Size (L / r) | Mass |
|---|---|---|---|---|---|---|
| Pelvis (root) | / | / | / | / | 0.86 / 0.23 m | 8 kg |
| Spine | 3 | -30 ~ 30° | -30 ~ 30° | -30 ~ 30° | 0.75 / 0.18 m | 18 kg |
| Left/right hip | 2 | -60 ~ 90° | -40 ~ 40° | / | 1.05 / 0.20 m | 14 kg |
| Left/right knee | 1 | -120 ~ 0° | / | / | 0.95 / 0.15 m | 7 kg |

**Table S3. Control performance summary for healthy and patient-adapted policies.** Mean values of speed-tracking accuracy and step-length accuracy across all healthy subjects and the model-assisted patient group ($n$ = 11). The healthy-average policy represents the population-level prior, while the patient-adapted policy denotes individualized fine-tuning results. Both maintain consistent control stability, with slightly higher mean accuracy after adaptation, reflecting alignment to patient-specific gait regimes without overfitting to individual variance.

| Policy | Speed Accuracy | Step-Length Accuracy |
|---|---|---|
| Healthy Motion Atlas | 90.04% | 89.94% |
| Atlas-to-Patient Adaptated Policy | 94.17% | 92.20% |

**Table S4. Quantitative training results across curriculum levels for the slope-ascent task.** Shown for a representative patient. Five curriculum levels (0–4) increase the ramp incline from flat ground to 1:12 (0, 1:30, 1:20, 1:15, 1:12; ramp length fixed). For each level, we report the training step interval, GAIL discriminator output, and the derived activation-signal imitation similarity score *S* (defined in Methods). These quantities are used as model-internal training diagnostics to monitor activation-pattern alignment during terrain-adaptive training. The results show that generated activation patterns remain highly consistent with the patient's reference pattern across slope difficulty levels, with small deviations at steeper slope conditions reflecting increased biomechanical complexity.

| **Curriculum Level** | **0** | **1** | **2** | **3** | **4** |
|---|---|---|---|---|---|
| Slope ratio | 0 | 1:30 | 1:20 | 1:15 | 1:12 |
| Training step interval | 0.7 M | 2.3 M | 4.7 M | 5.9 M | 9.4 M |
| GAIL discriminator output | 0.5007 | 0.5051 | 0.5040 | 0.5010 | 0.5015 |
| Activation-signal imitation similarity *S* | 99.86% | 98.98% | 99.20% | 99.80% | 99.70% |

**Table S5. Quantitative training results across curriculum levels in the stair-climbing task.** Shown for a representative patient. Five curriculum levels (0–4) correspond to increasing stair heights from 0 to 15 cm (0, 3.75, 7.50, 11.25, 15.00 cm). We report the training step interval, GAIL discriminator output, and activation-signal imitation similarity *S*. Reduced similarity at higher levels reflects greater task demands and biomechanical adaptation during stair-climbing.

| **Curriculum Level** | **0** | **1** | **2** | **3** | **4** |
|---|---|---|---|---|---|
| Step height | 0.00 cm | 3.75 cm | 7.50 cm | 11.25 cm | 15.00 cm |
| Training step interval | 11.3 M | 13.4 M | 26.1 M | 43.9 M | 57.3 M |
| GAIL discriminator output | 0.5023 | 0.5019 | 0.5849 | 0.5912 | 0.6014 |
| Activation-signal imitation similarity *S* | 99.54% | 99.62% | 83.02% | 81.76% | 79.72% |

**Table S6. Robustness evaluation of the data-physics hybrid control policy.** To assess the controller's resilience to external disturbances, impulsive forces were applied at the hip joint along forward (x-axis) and lateral (z-axis) directions until balance loss. Each perturbation was a 1-s pulse, with magnitude increased by 20 N per step, starting from 0 N, with 50 randomized trials conducted per condition. The hybrid control policy maintains stability under forward perturbations up to 980 N and lateral perturbations of 320–360 N on the affected and sound sides, respectively. Once the forward force exceeded 1,800 N or the lateral force approached 1,000 N, balance was lost. These results demonstrate that the learned controller achieves robust disturbance recovery, comparable to or exceeding prior benchmarks.

| **Direction of Perturbation** | **Applied Force Magnitude under Different Motion States / N** | | |
|---|---|---|---|
| | **Normal Walking** | **Moving in the Direction of the Force** | **Fall** |
| Forward | ≤ 980 | 980 ~ 1800 | ≥ 1800 |
| Lateral (affected side) | ≤ 320 | 320 ~ 1080 | ≥ 1080 |
| Lateral (sound side) | ≤ 360 | 360 ~ 920 | ≥ 920 |

**Table S7. Baseline comparison.** This table compares the proposed data–physics hybrid framework with a physics baseline with reference-tracking. Both models were trained on a 1:12 slope-climbing task under identical initialization and reward design. The hybrid framework, integrating behavior cloning (BC), generative adversarial imitation learning (GAIL), and curriculum-based adaptation, converged within 23 M environment steps versus ≈ 90 M for the baseline—a 74.4 % reduction in training time. It also achieved 4.73% higher joint-angle imitation consistency and 12.10% higher distal-joint (end-effector) positional imitation consistency, indicating faster convergence, stronger coordination, and more realistic motion than the physics-based reference-tracking baseline.

| Metric | Data–Physics Hybrid (Ours) | Physics with Reference-Tracking (DeepMimic) |
|---|---|---|
| Training steps to saturation | 23 M | 90 M |
| Cumulative reward | 280.1 | 270.7 |
| Joint-angle similarity | 73.02% | 69.72% |
| End-effector position similarity | 79.49% | 70.91% |

**Table S8. Joint profile PCC for terrain-adaptive motion prediction.** Values are mean ± SD across the model-assisted group (n = 11) for terrain-adaptive motion prediction in slope-ascent and stair-climbing tasks. Joint profile PCC was calculated between generated and measured joint-angle trajectories after time normalization. Values closer to 1 indicate stronger waveform-level agreement between predicted and measured joint profiles across the corresponding task conditions.

| Difficulty level | Slope Ascent | Stair Climbing |
|---|---|---|
| Level 1 | 0.93 ± 0.04 | 0.91 ± 0.04 |
| Level 2 | 0.92 ± 0.04 | 0.90 ± 0.04 |
| Level 3 | 0.91 ± 0.05 | 0.89 ± 0.05 |
| Level 4 | 0.90 ± 0.05 | 0.87 ± 0.06 |

**Table S9. Rehabilitation exposure and between-group statistical comparison.** Session-level records were compiled from all five participating sites across the 28-day rehabilitation program, during which participants received approximately three hours of supervised therapy per day. Values are mean ± SD unless otherwise indicated; percentages represent the proportion of total session time per exercise category. Total walking practice was calculated as the sum of overground walking, slope ascent, and stair climbing. p-values were calculated using independent-samples t-tests where applicable. Statistical significance was assessed using a two-sided α level of 0.05; $p < 0.05$ was considered statistically significant, whereas $p \geq 0.05$ was interpreted as indicating no statistically significant difference between groups. MA, model-assisted group.

| Variable | Model-Assisted (n = 11) mean ± SD | Control (n = 10) mean ± SD | p-value | Interpretation |
|---|---|---|---|---|
| Rehabilitation program duration (days) | 28.0 ± 0.0 | 28.0 ± 0.0 | — | Identical across groups |
| Daily therapist-contact time (min) | 180.2 ± 8.4 | 179.8 ± 8.1 | 0.91 | No significant difference |
| Total therapist-contact time (min) | 5046 ± 235 | 5034 ± 227 | 0.91 | No significant difference |
| Therapist clinical experience (yrs) | 8.7 ± 3.1 | 8.4 ± 3.0 | 0.82 | No significant difference |
| Balance and postural-control (%) | 13.0 ± 2.3 | 13.3 ± 2.6 | 0.78 | No significant difference |
| Lower-limb strengthening (%) | 20.0 ± 3.4 | 19.8 ± 2.9 | 0.89 | No significant difference |
| Range-of-motion and stretching (%) | 17.0 ± 2.8 | 16.9 ± 2.4 | 0.93 | No significant difference |
| **Total walking practice (%)** | 50.3 ± 2.1 | 49.7 ± 1.8 | 0.49 | No significant difference |
| Overground walking (%) | 22.3 ± 5.3 | 28.2 ± 2.0 | 0.004 | Lower overground walking time in MA group |
| Slope ascent practice (%) | 16.0 ± 6.3 | 12.5 ± 1.7 | 0.11 | Higher mean; greater variability consistent with individualized allocation |
| Stair climbing practice (%) | 12.0 ± 4.9 | 9.0 ± 1.7 | 0.08 | Higher mean; greater variability consistent with individualized allocation |

**Table S10. Patient cohort demographics.** A total of 21 Chinese post-stroke patients in the mid-recovery phase (1–6 months after neurosurgery) were enrolled at Beijing Tsinghua Changgung Hospital, Beijing Tongren Hospital, Shijiazhuang People's Hospital, Beijing University of Chinese Medicine, and the Fifth Medical Center of the Chinese PLA General Hospital (301 Hospital). All patients were able to stand and walk independently without assistive devices, and their group assignment, age, sex, height, weight, affected side, and time since stroke are summarized below.

| Subject ID | Group [a] | Age (years) | Sex | Height (cm) | Weight (kg) | Affected side | Time since stroke (months) | FMA-LE [b] | Stroke type [c] |
|---|---|---|---|---|---|---|---|---|---|
| S1 | MA | 23 | F | 162 | 65 | Left | 2.5 | 30 | IS |
| S2 | CT | 31 | M | 179 | 72 | Right | 5.0 | 25 | IS |
| S3 | MA | 47 | M | 176 | 67 | Right | 5.5 | 26 | IS |
| S4 | CT | 50 | M | 172 | 62 | Left | 1.5 | 26 | IS |
| S5 | CT | 54 | M | 169 | 68 | Right | 3.0 | 24 | HS |
| S6 | MA | 55 | M | 182 | 83 | Left | 1.5 | 25 | IS |
| S7 | CT | 56 | F | 157 | 55 | Right | 1.0 | 30 | IS |
| S8 | CT | 56 | M | 174 | 70 | Right | 4.0 | 25 | HS |
| S9 | MA | 58 | F | 163 | 54 | Left | 3.0 | 24 | IS |
| S10 | CT | 58 | M | 177 | 66 | Right | 5.0 | 27 | HS |
| S11 | MA | 60 | M | 178 | 78 | Right | 2.5 | 26 | IS |
| S12 | MA | 61 | M | 180 | 80 | Left | 2.0 | 27 | IS |
| S13 | CT | 63 | F | 160 | 51 | Left | 3.0 | 22 | IS |
| S14 | MA | 64 | F | 166 | 58 | Left | 1.0 | 33 | IS |
| S15 | CT | 64 | M | 169 | 62 | Left | 2.5 | 26 | IS |
| S16 | MA | 65 | M | 181 | 74 | Right | 3.5 | 25 | HS |
| S17 | MA | 67 | F | 160 | 55 | Left | 2.0 | 26 | IS |
| S18 | CT | 69 | M | 169 | 80 | Left | 3.0 | 25 | IS |
| S19 | MA | 69 | M | 178 | 69 | Right | 5.0 | 26 | HS |
| S20 | CT | 70 | M | 174 | 70 | Left | 3.5 | 26 | HS |
| S21 | MA | 71 | M | 170 | 58 | Right | 2.0 | 22 | HS |

[a] MA: model-assisted group; CT: control group. [b] Baseline lower-extremity motor impairment assessed by the Fugl–Meyer Assessment (FMA-LE) prior to study initiation. [c] IS: Ischemic stroke; HS: hemorrhagic stroke.

**Table S11. Dataset composition.** The dataset integrates two motion sources—custom-captured sequences and public mocap recordings—covering both healthy participants and stroke patients across a range of walking conditions. For each cohort, we report the number of clips, within-sequence consistency (quantifying cycle-to-cycle agreement of step length and walking speed), average sequence duration, and sampling rate. Higher consistency values indicate more stable and regular gait cycles, reflecting greater motor control, whereas slightly lower scores in patient data capture expected asymmetry and variability. For each stroke patient, one representative walking clip is selected near the subject's median speed and step length, providing a realistic depiction of typical gait performance. For healthy participants, multiple walking trials at different natural speeds and step lengths were available; representative clips were selected from these trials.

| Data Source | Samples | Number of Clips | Consistency | Mean Duration/s | Sampling Rate/Hz |
|---|---|---|---|---|---|
| Custom capture | Healthy people | 44 | 92.55% | 0.93 | 90 |
| | Stroke patients | 21 | 82.21% | 0.94 | 90 |
| Public MOCAP | Healthy people | 118 | 97.11% | 1.16 | 120 |

**Table S12. Hyperparameter configuration of the data–physics hybrid model.** Key settings for the policy, value, and GAIL discriminator networks (hidden-layer widths, learning rates) and training weights (behavior-cloning action-blending; external/internal reward weighting).

| Model Hyperparameter | Value |
|---|---|
| Hidden layers / nodes in policy network | 3/256 |
| Hidden layers / nodes in value network | 2/128 |
| Hidden layers / nodes in GAIL discriminator | 2/128 |
| Loss blending weight of BC | 0.2 → 0 (linear, by 5 M steps) |
| External / Internal reward weight | 0.9/0.1 |
| PPO learning rate | $2\times10^{-5}\rightarrow 0$ (linear) |
| GAIL learning rate | $1\times10^{-5}\rightarrow 0$ (linear) |
| Activation (all hidden layers) | ReLU |
| Simulation / control frequency | 30 Hz |
| Rollout size per PPO update | 204,800 samples |
| Minibatch size | 2,048 |
| Discount factor ($\gamma$) | 0.995 |
| GAE ($\lambda$) | 0.95 |
| PPO clip coefficient ($\varepsilon$) | 0.2 |

**Movie S1. Animated walkthrough of the framework and experimental results**.